\documentclass[11pt]{article}
\usepackage[super,sort&compress,comma]{natbib} 

\usepackage{graphicx} %eps figures can be used instead
\usepackage{lastpage}
\usepackage{color}
\usepackage[format=plain,justification=raggedright,singlelinecheck=false,font=small,labelfont=bf,labelsep=space]{caption} 

\usepackage{verbatim}
\usepackage[table]{xcolor}

\title{Understanding corrosion inhibition with van der Waals DFT methods: the case of benzotriazole}
\author{Chiara Gattinoni and Angelos Michaelides\\Thomas Young Centre, London Centre for Nanotechnology and Department of Chemistry\\ University College London\\17-19 Gordon Street, London, WC1H 0AH, UK}
\date{}

\begin{document}
\maketitle

\begin{abstract}

The corrosion of materials is an undesirable and costly process affecting many areas of technology and everyday life.
As such, considerable effort has gone into understanding and preventing it.
Organic molecule based coatings can in certain circumstances act as effective corrosion inhibitors.  
Although they have been used to great effect for more than sixty years, how they function at the atomic-level is still a matter of debate.
In this work, computer simulation approaches based on density functional theory are used to investigate benzotriazole (BTAH), one of the most widely used and studied corrosion inhibitors for copper.
In particular, the structures formed by protonated and deprotonated BTAH molecules on Cu(111) have been determined and linked to their inhibiting properties.
It is found that hydrogen bonding, van der Waals interactions and steric repulsions all contribute in shaping how BTAH molecules adsorb, with flat-lying structures preferred at low coverage and upright configurations preferred at high coverage.
The interaction of the dehydrogenated benzotriazole molecule (BTA) with the copper surface is instead dominated by strong chemisorption via the azole moiety with the aid of copper adatoms.
Structures of dimers or chains are found to be the most stable structures at all coverages, in good agreement with scanning tunnelling microscopy results.
Benzotriazole thus shows a complex phase behaviour in which van der Waals forces play an important role and which depends on coverage and on its protonation state and all of these factors feasibly contribute to its effectiveness as a corrosion inhibitor.
\end{abstract}

\section{Introduction}
\label{intro}
The oxidation and corrosion of metals affects many areas of industry, technology, and everyday life.
Indeed, it plays a critical role in the failure of metal parts in engineering, electrochemical and catalytic devices, as well as of metal construction parts such as piping and roofing, to name but a few.
The atomistic details of the processes of metal corrosion are still unclear, however an effective solution to the problem of corrosion has been found in the use of organic molecules as corrosion inhibitors, which work by interacting with metal/oxide surfaces and forming a protective film~\cite{breston, antonijevic}.
Widely-used organic corrosion inhibition systems, whose effectiveness has been empirically verified, are amines or zinc dithiophosphates on steel and benzotriazole (BTAH~\cite{btah_patent}) on copper.
However, the chemical details of the inhibiting action and the structure of the protective layer these molecules form against metal surfaces are still unresolved.
In this work, state-of-the-art computational methods are applied to the BTAH/Cu copper system, which is the most studied both experimentally~\cite{finsgar} and computationally~\cite{kokalj_lang, jiang_2003, peljhan_pccp_2011, peljhan_jpcc_2014, kokalj_jacs_2010, hakkinen}.
The aim is to identify the structures BTAH forms on the copper surface and obtain insights into how they affect its function as a corrosion inhibitor.

Most experimental studies on organic corrosion inhibitors are macroscopic studies involving the immersion of a metal specimen in a solution containing corrosive agents and organic inhibitors.
Information about inhibition can be gathered during sample immersion, monitoring the evolution of the material using electrochemical or spectroscopic techniques (see \emph{e.g.} Refs.~\cite{dugdale, roberts_1974, rubim}) or after immersion, by measuring the mass change of the specimen or the amount of metal solvated in the solution (see \emph{e.g.} Ref.~\cite{rathgeber}).
These studies reveal insight into the effectiveness of different chemicals.
An alternative approach is to examine well-defined copper single-crystal surfaces under ultra-high vacuum (UHV) using X-ray absorption spectroscopy (XAS) and scanning tunnelling microscopy (STM).
These techniques, used mainly by the surface science community, reveal structural details of the adsorbed system.
The inhibitor molecules can be deposited onto the surface either from gas phase or via an aqueous solution (which is subsequently evaporated).
Discrepancies are seen for different experimental conditions, with zigzag  structures of flatly-adsorbed molecules having been postulated~\cite{dugdale, roberts_1974, rubim} (for molecules deposited from aqueous solutions), as well as a structure of vertically-adsorbed BTAs~\cite{fang_olson} (for evaporated molecules).
In this work we only compare our results with experiments performed on evaporated BTAH molecules on copper substrates in UHV, since the computational work presented here consists of molecules adsorbed on a Cu(111) surface in vacuum conditions.
Previous studies have shown that BTAH deprotonates when adsorbed on copper surfaces, revealing that it can deprotonate not only by interacting with high-pH environments (its pK$_a$ is $8.4$ at $25^{\circ}$ C) but also through its interaction with the surface.
BTA was seen to form nearly-upright organometallic surface complexes involving bonds between the azole nitrogen atoms and copper adatoms~\cite{walsh, grillo_nano_2013, grillo, fang_olson}.
In particular STM studies showed the formation of strings of stacked dimers at low coverage and of more complex structures, also composed of BTA-Cu$_{\mathrm{ad}}$-BTA dimers, at high coverage.

A number of computational studies, using density functional theory (DFT), have looked at the stability of BTAH and BTA structures on copper.
Benzotriazole is a challenging molecule to study with DFT as it combines a strongly electronegative azole moiety, which preferably interacts with the surface through chemisorption, with a benzene-like ring which can interact with the surface via van der Waals (vdW) forces.
Traditionally, vdW dispersion forces have been a challenge for DFT methods and are not accounted for in the most widely used exchange-correlation functionals, such as Perdew-Burke-Ernzerhof (PBE)~\cite{pbe}, that have been used in most previous studies of BTAH.
Thus the role of vdW forces in this type of system is still largely unexplored.
Isolated BTAH was found, in calculations using PBE, to chemisorb weakly to a Cu(111) surface in an upright geometry, forming two N-Cu bonds via the triazole moiety~\cite{kokalj_lang, jiang_2003}.
Physisorption, with the molecule lying flat onto the surface, was found to yield a very small binding energy ($\sim 0.1$ eV) with PBE~\cite{jiang_2003}.
The addition of an empirical van der Waals correction lead instead to a much stronger bond ($\sim 0.72$ eV)~\cite{kokalj_lang}.
At higher coverages~\cite{kokalj_jacs_2010}, BTAH was found to form hydrogen-bonded (HB) chains with the molecules lying parallel to the surface.
The dehydrogenated BTA molecule was found to bind strongly to the surface in an upright configuration~\cite{kokalj_lang} when isolated, and to form Cu-bonded organometallic complexes of vertical or tilted molecules at higher coverages.
However, the structure of the organometallic BTA-Cu$_{\mathrm{ad}}$ complexes is still debated.
Chen and Hakkinen~\cite{hakkinen} found that deprotonated BTA-Cu$_{\mathrm{ad}}$-BTA dimers are more stable than [BTA-Cu$_{\mathrm{ad}}$]$_n$ chains thus agreeing with the experimental results of Grillo \emph{et al.}~\cite{grillo}, and in disagreement with the DFT results of Kokalj \emph{et al.}~\cite{peljhan_jpcc_2014}.

Here we report the most extensive DFT study performed to date and we explore in depth the importance of using vdW approaches in the study of adsorbate-surface interactions.
Intramolecular interactions and coverage effects are also examined in detail; they are found to be particularly relevant for the systems formed by fully protonated BTAHs, with hydrogen bonding dominating in the low-coverage structures and vdW and electrostatic forces at high coverage.
Moreover, the energetics of the dissociation process and of the formation of complex structures with Cu adatoms were investigated and linked to experimental conditions.
The obtained structures and adsorption energy trends are comparable to the experimental results in UHV and a link with the effectiveness of the molecule as a corrosion inhibitor is discussed.

The remainder of the paper is organised as follows.
The computational methodology and set-up is presented in the next section (Sec.~\ref{method}), followed by the results for the protonated (Sec.~\ref{sub:btah_cu111}) and deprotonated (Sec.~\ref{sub:bta_cu111}) molecules adsorbed on Cu(111).
Finally, a discussion and conclusions are presented in Sec.~\ref{conclusions}.

\section{Methodology and Computational Setup}
\label{method}

Calculations of inhibitor molecules adsorbed on copper surfaces were performed by means of DFT using the VASP code~\cite{vasp_1, vasp_2, vasp_3, vasp_4}.
In the present work most results have been obtained with the optB86b-vdW functional, a modified version of the non-local vdW density functional~\cite{dion_prl_2004} which explicitly accounts for dispersion based on the electron density.
Indeed, optB86b-vdW has been shown to perform best in comparison to experiment for several adsorption problems~\cite{carrasco_prl, klimes_review}, including the (relevant for this work) adsorption of benzene on Cu(111)~\cite{carrasco}.
A number of other functionals were also considered for testing and comparison with previous results: PBE and a number of vdW-inclusive functionals (vdW-DF~\cite{dion_prl_2004}, optB86b-vdW~\cite{klimes}, PBE-D2~\cite{grimme} and PBE-TS~\cite{tkatchenko}).
For all functionals the calculated value for the lattice constant $a$ and of the bulk modulus $B$ are within $10 \%$ of the experimental value (see Table~\ref{table:tests}), and in good agreement with previous theoretical results~\cite{klimes}.
\begin{table}[h]
 \begin{tabular}{|c|c|c|}
   \hline
      & $a$ [\AA] & $B$ [GPa] \\
   \hline
   PBE &  3.64 & 137 \\
   vdW-DF & 3.71 & 113 \\
   optB86b-vdW & 3.60 & 148 \\
   PBE-D2 & 3.57 & 148 \\
   PBE-TS & 3.55 & 168 \\
   \hline
   Expt. & 3.49 & 137 \\
   \hline
 \end{tabular}
 \caption{Lattice constants and bulk moduli for all the exchange-correlation functionals used in this work. All the values are in good agreement with the experimental values corrected for the zero-point anharmonic expansion~\cite{haas} and with previous computational results~\cite{klimes}.}
 \label{table:tests}
\end{table}

Slabs of 3-6 layers were tested with optB86b-vdW (the functional chosen for the calculations of the BTAH/Cu systems), and the surface energy was seen to converge at a 4-layer thickness to $\gamma^{\mathrm{optB86b-vdW}}_{\mathrm{Cu(111)}}=0.102$ eV/\AA$^{2}$, in good agreement with the experimental value~\cite{haas} ($\gamma^{\mathrm{exp}}_{\mathrm{Cu(111)}}=0.112$ eV/\AA$^{2}$).
In adsorption calculations the atoms in the bottom layer of the slab were kept fixed at the bulk positions and a vacuum of $\sim20$ \AA\ in the non-periodic direction was added to prevent interactions between molecules in neighbouring cells.
All calculations were performed using the PAW method~\cite{paw, paw_vasp} and a kinetic energy cutoff on the plane wave basis set of $400$ eV. 
The Brillouin-zone integration was performed using Monkhorst-Pack grids of ($12\times12\times12$) \textbf{k}-points for the conventional unit cell of copper bulk and of ($12\times12\times1$) \textbf{k}-points for a $1 \times 1$ Cu(111) surface with the vacuum in the $z$ direction.
A dipole correction as implemented in the VASP code~\cite{neugebauer_prb_1992, makov_prb_1995} was added in all cases, and calculations of the dissociated molecule (into BTA and H) in the gas-phase were spin-polarized.
The inhibitor (BTAH and BTA) molecules were adsorbed onto the surface at a variety of coverages, between 1/16 to 1/4 of a monolayer (ML), where 1 ML corresponds to 1 adsorbed molecule per surface copper atom.
Experimental studies looking at the sub-ML regime were performed at 1/16 ML~\cite{grillo} and at 1/3 ML~\cite{walsh}, close to the coverages studied in this work.
A variety of hexagonal and orthorhombic cells were used and a large number of initial configurations have been chosen and optimised, in order to explore phase space as comprehensively as possible.

The adsorption energies, $E_{\mathrm{ads}}$, were calculated as follows:
\begin{equation}
 E_{\mathrm{ads}}=(E_{\mathrm{sys}}-E_{\mathrm{slab}}-N_{\mathrm{mol}}\times E_{\mathrm{mol}})/N_{\mathrm{mol}},
 \label{eq1}
\end{equation}
where $E_{\mathrm{sys}}$ is the total energy of the adsorbed complex, $E_{\mathrm{slab}}$ is the total energy of the substrate (a Cu(111) slab with or without adatoms), $E_{\mathrm{mol}}$ is the total energy of the isolated molecule and $N_{\mathrm{mol}}$ is the number of molecules in each cell.
A negative value of $E_{\mathrm{ads}}$ indicates a stable adsorbed structure.

The formation of complex systems, such as deprotonated BTAH on a copper surface with defects, includes endothermic (bond breaking and defect formation) and exothermic (adsorption of the molecules on the surface) processes.
In order to account for all these energy contributions, an energy of formation, E$_{\mathrm{form}}$, is defined here and used as an indicator of the likelihood of formation of the system.
The following two stoichiometric equations are relevant to the formation of a system of adsorbed BTA molecules.
In the first one, $n$ BTAH molecules dissociate and adsorb on a Cu(111) surface where $m$ copper adatoms Cu$_\mathrm{ad}$ are already present, to form the $n \mathrm{BTA}/\mathrm{Cu}\mathrm{(111)}/m \mathrm{Cu}_\mathrm{ad}$ organometallic complex:
\begin{equation}
n \mathrm{BTAH}+\mathrm{Cu}\mathrm{(111)}/m \mathrm{Cu}_{\mathrm{ad}} \rightarrow n \mathrm{BTA}/\mathrm{Cu}\mathrm{(111)}/m \mathrm{Cu}_\mathrm{ad}+\mathrm{H}_{\mathrm{sys}}.
 \label{reaction1}
\end{equation}
The second process is similar, however the copper surface is assumed to initially be atomically flat Cu(111) and $m$ copper adatoms Cu$_\mathrm{ad}$ are formed from the bulk:
\begin{equation}
n \mathrm{BTAH}+\mathrm{Cu}\mathrm{(111)} + m \mathrm{Cu}_{\mathrm{bulk}} \rightarrow n \mathrm{BTA}/\mathrm{Cu}\mathrm{(111)}/m \mathrm{Cu}_\mathrm{ad}+\mathrm{H}_{\mathrm{sys}}.
 \label{reaction2}
\end{equation}
In Eqs.~\ref{reaction1}-\ref{reaction2}, the term H$_{\mathrm{sys}}$ indicates either of two possible scenarios for the dissociated hydrogen atoms: they can either adsorb onto the copper surface or combine to form gas-phase hydrogen molecules.
The formation energy, E$_{\mathrm{form}}$, is defined as the difference in total energy between the final (right-hand side of Eqs.~\ref{reaction1}-\ref{reaction2}) and the initial system (left-hand-side of Eqs.~\ref{reaction1}-\ref{reaction2}).
Note that the adsorption energy E$_{\mathrm{ads}}$ of BTAH on clean Cu(111) can be considered as a formation energy obtained from the stochiometric equation:
\begin{equation}
n \mathrm{BTAH}+\mathrm{Cu}\mathrm{(111)}  \rightarrow n \mathrm{BTAH}/\mathrm{Cu}\mathrm{(111)}.
 \label{reaction3}
\end{equation}

For certain overlayer structures, energy decompositions have been performed in order to estimate the strength of hydrogen bonds, of non-HB intermolecular interactions and of adsorbate-substrate bonding.
These are performed by taking the optimised system and performing single point calculations of the individual system components.
The strength of a hydrogen bond, $\mathrm{E}_{\mathrm{HB}}$, is obtained by performing single point calculations for isolated (gas phase) HB chains or dimers and for each gas phase BTAH molecule:
\begin{equation}
\mathrm{E}_{\mathrm{HB}}=(\mathrm{E}_{\mathrm{N} \times \mathrm{BTAH}}^{\mathrm{gas}}-\sum_{\mathrm{N}} \mathrm{E}_{\mathrm{BTAH}})/\mathrm{N},
\label{decomp1}
\end{equation}
where $\mathrm{E}_{\mathrm{N} \times \mathrm{BTAH}}^{\mathrm{gas}}$ is the total energy in gas phase of the HB system of N$\times$BTAH molecules, and $\mathrm{E}_{\mathrm{BTAH}}$ is the total energy of each BTAH molecule in the chain.
Similarly, non-HB interactions are estimated using:
\begin{equation}
\mathrm{E}_{\mathrm{non-HB}}=(\mathrm{E}_{\mathrm{N \times BTAH}}^{\mathrm{unit}}-\sum_{\mathrm{N}} \mathrm{E}_{\mathrm{BTAH}}-\mathrm{E}_{\mathrm{HB}})/\mathrm{N},
\label{decomp3}
\end{equation}
where $\mathrm{E}_{\mathrm{N \times BTAH}}^{\mathrm{unit}}$ is the total energy obtained by removing the copper surface from the unit cell and performing a single point calculation, $\mathrm{E}_{\mathrm{BTAH}}$ is the total energy of each BTAH molecule in the system and $\mathrm{E}_{\mathrm{HB}}$ is the strength of the HB interactions (if any) calculated using Eq.~\ref{decomp1}.
The interaction energy of the molecules with the substrate is obtained using:
\begin{equation}
\mathrm{E}_{\mathrm{mol/Cu(111)}}=(\mathrm{E}_{\mathrm{sys}}-\mathrm{E}_{\mathrm{N \times BTAH}}^{\mathrm{unit}} -\mathrm{E}_{\mathrm{Cu(111)}})/\mathrm{N},
\label{decomp2}
\end{equation}
where $\mathrm{E}_{\mathrm{sys}}$ is the total energy of the adsorbed system, $\mathrm{E}_{\mathrm{N \times BTAH}}^{\mathrm{unit}}$ and $\mathrm{E}_{\mathrm{Cu(111)}}$ the total energies obtained by performing a single point calculation after removing, respectively, the surface and the molecules from the unit cell.
Energy decomposition schemes entail a certain level of arbitrariness~\cite{michaelides_prb_2004}, however they are also useful in providing a semi-quantitative understanding of the relative importance of hydrogen bonding versus adsorbate-substrate bonding.

The nature of the bonding between the adsorbates and substrate was investigated by looking at electron density difference plots.
They were constructed by subtracting from the charge density of an optimised adsorbed system the charge density of the adsorbates and substrate.
As for the energy decompositions, single point calculations are performed on the components \emph{i.e.} they are kept to the structure they assume in the adsorption system.

\section{BTAH on Cu(111)}
\label{sub:btah_cu111}

The adsorption of benzotriazole on Cu(111) was studied first for a variety of coverages.
The Cu(111) surface was chosen since it has the lowest surface energy among the copper surfaces, and therefore it would be the most prevalent surface even on a copper polycrystalline sample.
Moreover, it is the most widely studied surface experimentally and computationally.
 
Although experiments in vacuum have shown that BTAH in contact with a copper surface deprotonates~\cite{grillo, grillo_nano_2013, walsh}, it is still useful to look at the adsorption of the fully protonated molecules for at least two reasons. 
First, the drive to deprotonation can depend on the environment and in acidic environments the molecule can be expected to be protonated.
Indeed, a strong dependence on pH has been found on the effectiveness of benzotriazole as a corrosion inhibitor~\cite{musiani_jec_1987} (with worse inhibition in acidic media), suggesting that the molecular structure changes with pH and, thus, with the BTAH/BTA ratio present in the system.
Second, the BTAH/Cu(111) system is well-studied and it provides the possibility of validating the results obtained here with previous work.

A large number of DFT calculations (summarized in Fig.~\ref{fig:alt_fig_1}-\ref{fig:int_energies}) was performed for BTAH on Cu(111) considering molecules at different adsorption sites, different orientations and different coverages.
In this section we are going to look first at the behaviour of isolated molecules (Sec.~\ref{isolated}), and subsequently at supramolecular complexes (Sec.~\ref{supramolecular}).
In particular, we are going to highlight the role of hydrogen bonding, van der Waals dispersion and chemical bonding in the coverage dependence of BTAH on Cu(111).

\subsection{Isolated BTAH on Cu(111)}
\label{isolated}
\begin{figure}[h]
\centering
 \includegraphics[width=1.0\textwidth]{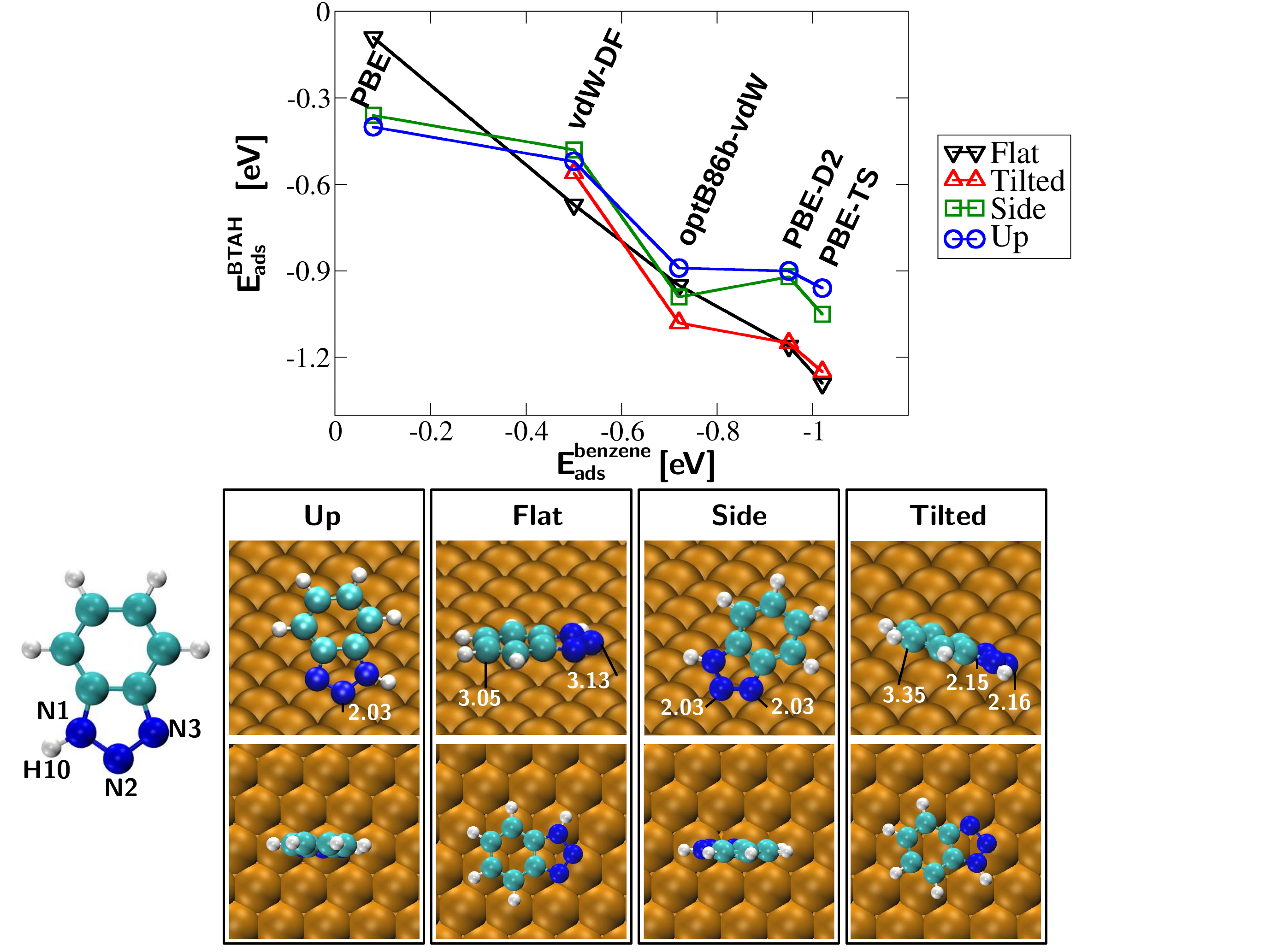}
 \caption{Top: adsorption energies for the Flat, Tilted, Side and Up structures calculated with five different functionals. They are plotted as a function of the adsorption energy of benzene on Cu(111), calculated with the corresponding functional. The adsorption energies of benzene on Cu(111) are taken from Refs.~\cite{klimes, kokalj_jacs_2010}. Bottom: structures of the four lowest energy configurations for a single BTAH adsorbed on Cu(111). Selected bond distances are given in \AA. The molecule and labelling scheme are shown on the left.}
 \label{fig:alt_fig_1}
\end{figure}

\begin{figure}[h]
\centering
 \includegraphics[width=1.0\textwidth]{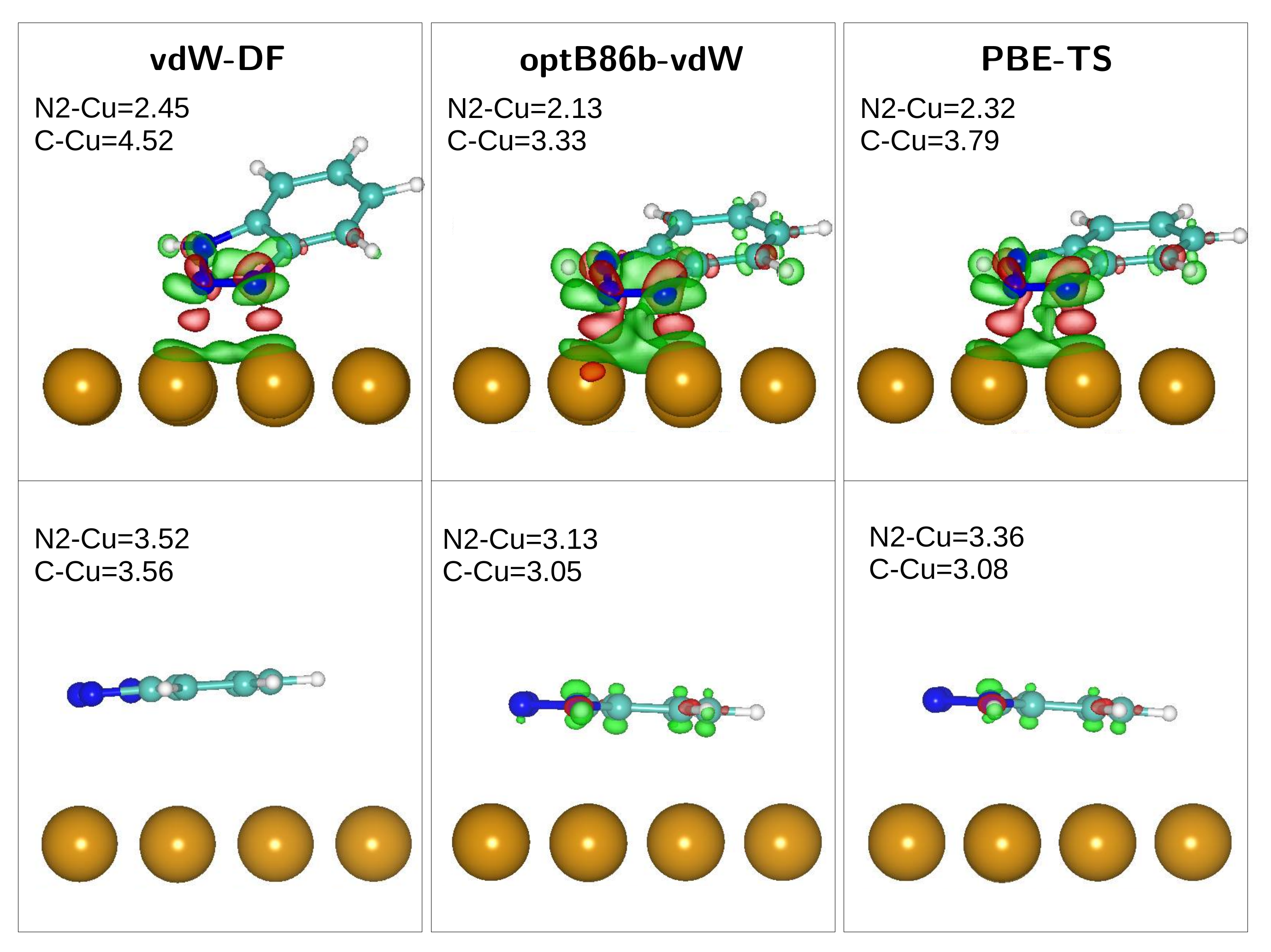}
 \caption{Electron density difference plots for the `Tilted' (top row) and `Flat' (bottom row) BTAH/Cu structures using three representative vdW-inclusive functionals. Green represents region of charge density depletion and red regions of accumulation. The isosurface level is $0.002$ $e/a_0^3$ for all structures. The distance between the azole N2 and the surface and between the centre of the benzene-like ring and the surface are given in \AA .}
 \label{fig:charge_dens}
\end{figure}

The most stable adsorption configurations for a single BTAH on Cu(111) at low coverage (1/16 ML) were established using both PBE and the vdW functionals mentiond in Sec.~\ref{method}. 
Four stable structures were obtained and they are shown in Fig.~\ref{fig:alt_fig_1}.
Two upright configurations were found: the `Up' structure, which binds to the surface via the N2 triazole atom and the `Side' structure which binds via the N2 and N2 triazole atoms.
A physisorbed flat configuration (`Flat' in Fig.~\ref{fig:alt_fig_1}) was also seen.
Finally, a not-previously seen tilted configuration (`Tilted' in Fig.~\ref{fig:alt_fig_1}) which combines chemisorption through two N-Cu bonds (as shown in Fig.~\ref{fig:charge_dens}), with physisorption through $\pi$-bonding via the benzene-like ring, was identified.

The comparison of their adsorption energies obtained with different functionals highlights the importance of choosing the most suitable methodology for the system under study.
Indeed, the graph in Fig.~\ref{fig:alt_fig_1} shows that the relative adsorption energies of the four structures changes when they are optimised with different functionals.
For example, the least stable structure with the PBE functional is the `Flat' configuration, since PBE is unable to describe physisorption.
This is however one of the most stable structures for all the other functionals.
The adsorption energy of the flat structure with PBE is $E_{\mathrm{ads}}=-0.10$ eV/mol, consistent with Ref.~\cite{jiang_ss}.
It is also in line with PBE calculations of benzene on Cu(111) ($E_{\mathrm{ads}}^{\mathrm{benzene}}=-0.08$ eV/mol)~\cite{klimes} which also showed almost no interaction of the molecule with the surface.
When dispersion interactions are accounted for, the stability of the system increases.
In addition, we find that the adsorption energy of the `Flat' structure is linearly related to that of the benzene ring (Fig.~\ref{fig:alt_fig_1}), and it was found to be between $E_{\mathrm{ads}}^{\mathrm{vdW-DF}}=-0.66$ to $E_{\mathrm{ads}}^{\mathrm{PBE-TS}}=-1.29$.
The nature of the interaction of the molecule with the surface, \emph{i.e.} physisorption, was confirmed by looking at electron density difference plots, shown, for selected functionals, in Fig.~\ref{fig:charge_dens}.
No interaction between the surface and the molecule is observed via the triazole ring, and only a small amount of charge rearrangement is observed for functionals which yield a stronger binding of benzene to Cu(111) (PBE-TS and optB86b-vdW).
The molecule-surface distance, when optimizing with functional which underbind the benzene ring (PBE and vdW-DF), is constant across the molecules ($3.55-3.65$ \AA ).
When functionals which strongly bind the benzene ring are used, a small amount of tilting is observed, in good agreement with Ref.~\cite{atodiresei}.

The most stable configurations with PBE are the upright chemisorbed configurations (`Up' and `Side' in Fig.~\ref{fig:alt_fig_1}).
Their PBE adsorption energies ($E_{\mathrm{ads}}=-0.40$  eV/mol for `Up' and $E_{\mathrm{ads}}=-0.36$ eV/mol for `Side') were found to be in excellent agreement with Ref.~\cite{kokalj_jacs_2010}.
An inversion of the energy trend between these two structures, \emph{i.e.} $E_{\mathrm{ads}}^{\mathrm{Side}}< E_{\mathrm{ads}}^{\mathrm{Up}}$, can be seen for optB86b-vdW, PBE-D2 and PBE-TS, but not for PBE-D2, thus showing that the use of different van der Waals interactions can also affect results involving chemisorption.

The inclusion of dispersion interactions in the optimisation of all the structures treated in this work allowed for the identification of the tilted structure (`Tilted' in Fig~\ref{fig:alt_fig_1}).
This was found to be the most stable configuration using optB86b-vdW (E$_{\mathrm{ads}}=-1.07$ eV) and degenerate to the `Flat' structure with PBE-D2 (E$_{\mathrm{ads}}=-1.15$ eV) and PBE-TS (E$_{\mathrm{ads}}=-1.29$ eV).
It is however unstable when optimised with PBE (the `Flat' configuration is recovered) and unfavourable with vdW-DF, the adsorption energy being $\sim 0.1$ eV/mol higher than for the `Flat' configuration.
This result is in good agreement with previous work on functionalized benzene rings with PBE and vdW-inclusive functionals: it was seen that the inclusion of dispersion interactions is the driving element behind the chemisorption of molecules which would otherwise weakly physisorb using PBE~\cite{atodiresei, liu_nc_2013}.
From Fig.~\ref{fig:charge_dens} it can be clearly seen that there is substantial rearrangement of the charge density at the azole end of the molecule and the Cu surface atoms to which it is bonded.
This is indicative of the formation of a weak chemical bond between adsorbate and substrate.
Therefore, from the results obtained for the `Flat' and `Tilted' configuration with PBE and the vdW-inclusive functionals it is clear that dispersion stabilises these two configurations by bringing the molecule closer to the surface.
This reinforces the chemical interaction of the molecule to the surface via the azole group, in the case of the `Tilted' structure and via $\pi$ bonding of the benzene ring, in the case of the `Flat' structure.

It can be noticed from the graph in Fig.~\ref{fig:alt_fig_1} that the difference between the absolute adsorption energies obtained  with the functional yielding the weakest adsorption energies (PBE) and the strongest (PBE-TS) is $\sim0.9$ eV/mol.
In general, absolute adsorption energies will inevitably vary between exchange-correlation functionals~\cite{klimes_review}. 
However, studies of benzene on Cu(111) have shown that the optB86b-vdW functional yields a binding energy in very good agreement with experiment, with the PBE-D2 and PBE-TS functionals overbinding and vdW-DF underbinding~\cite{carrasco}.
This makes optB86b-vdW the best candidate to study the relatively similar BTAH, and it has thus been chosen to obtain the results presented in the next sections.

\subsection{Coverage dependence of BTAH on Cu(111)}
\label{supramolecular}
\begin{figure}[h]
\centering
 \includegraphics[width=1.0\textwidth]{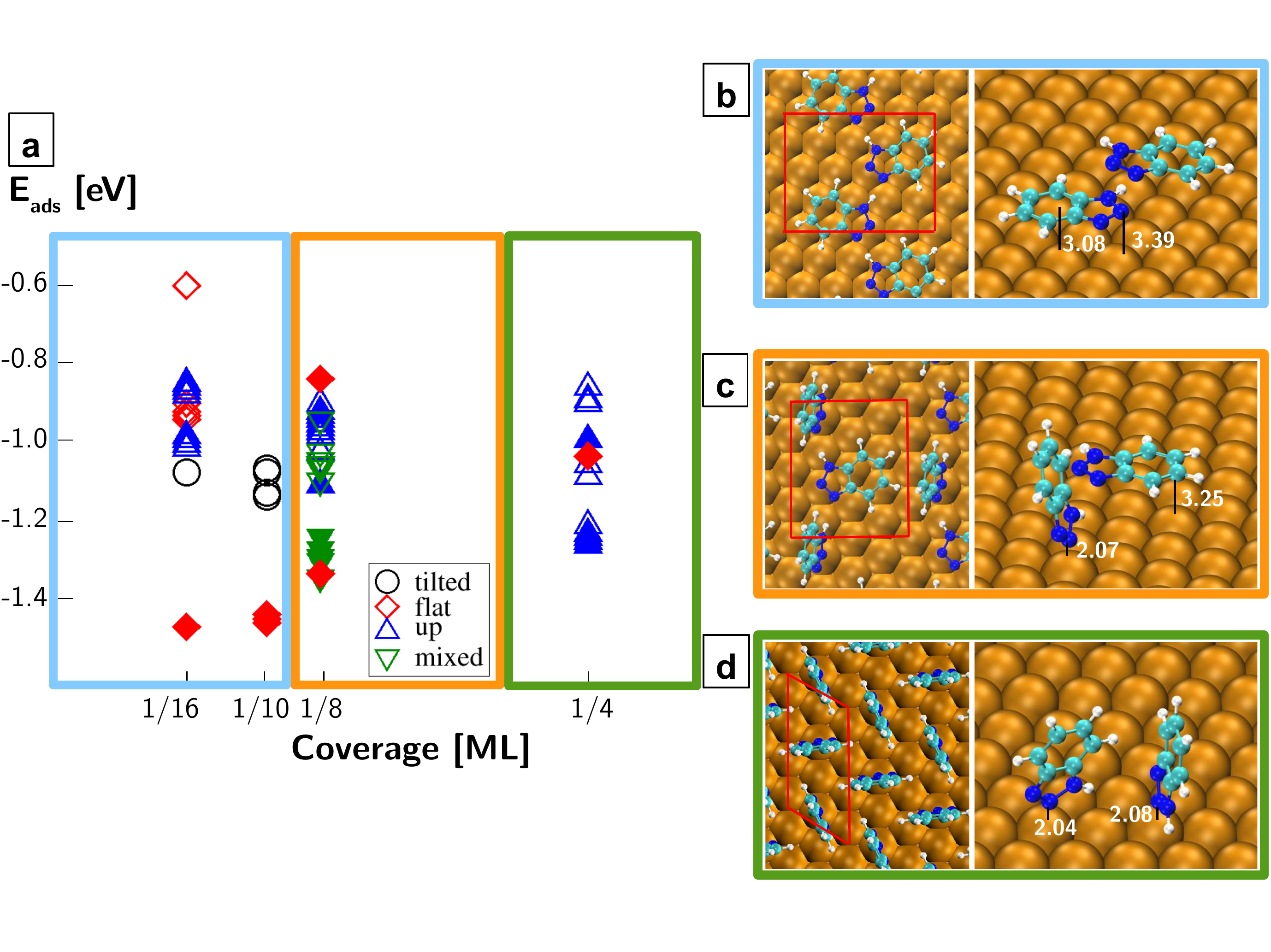}
 \caption{a) Adsorption energy of BTAH on Cu(111) as a function of coverage. Each point corresponds to a fully optimised adsorption structure. The filled symbols indicate a system where hydrogen bonding is present. The three coloured boxes define the three observed regimes: light blue for the physisorbed regime at low coverage, orange for the mixed adsorption regime at intermediate coverage and green for the chemisorption regime at the highest coverage. b-d) Top and side view of the most stable BTAH/Cu(111) structure for each regime. In the top view the red box indicates the unit cell and periodic images are shown. In the side view the periodic images are not shown for clarity. Selected distances are given in \AA . b) Physisorbed regime: BTAH forms HB chains of flat molecules on Cu(111). c) 1/8 ML coverage: mixed arrangement of HB flat and upright structures. d) 1/4 ML coverage: representative low-energy structure with vertically adsorbed molecules at a $60^{\circ}$ angle with respect to one another.}
 \label{fig:str_cov}
\end{figure}
Having established the most stable configurations for the isolated BTAH molecule on Cu(111), the stability of an extensive set of structures was evaluated with the optB86b-vdW functional as a function of coverage.
The results, for 1/16 ML, 1/10 ML, 1/8 ML and 1/4 ML coverage, are shown in Fig.~\ref{fig:str_cov}.
Each point of Fig.~\ref{fig:str_cov}a corresponds to a fully optimized adsorption structure for BTAH on Cu(111).
Three regimes of adsorption can be observed: at low coverage (up to 1/10 ML), hydrogen-bonded physisorbed structures were found to be dominant, at intermediate coverage (1/8 ML) an adsorption mode consisting of a mixture of physisorption and chemisorption is found, and at high coverage (1/4 ML) pure chemisorption is favoured.
Hydrogen bonding is present in all low-energy configurations.
It is dominant at low-coverage and it competes with other forms of intermolecular interaction at high coverage.

\begin{figure}[h]
\centering
 \includegraphics[width=1.0\textwidth]{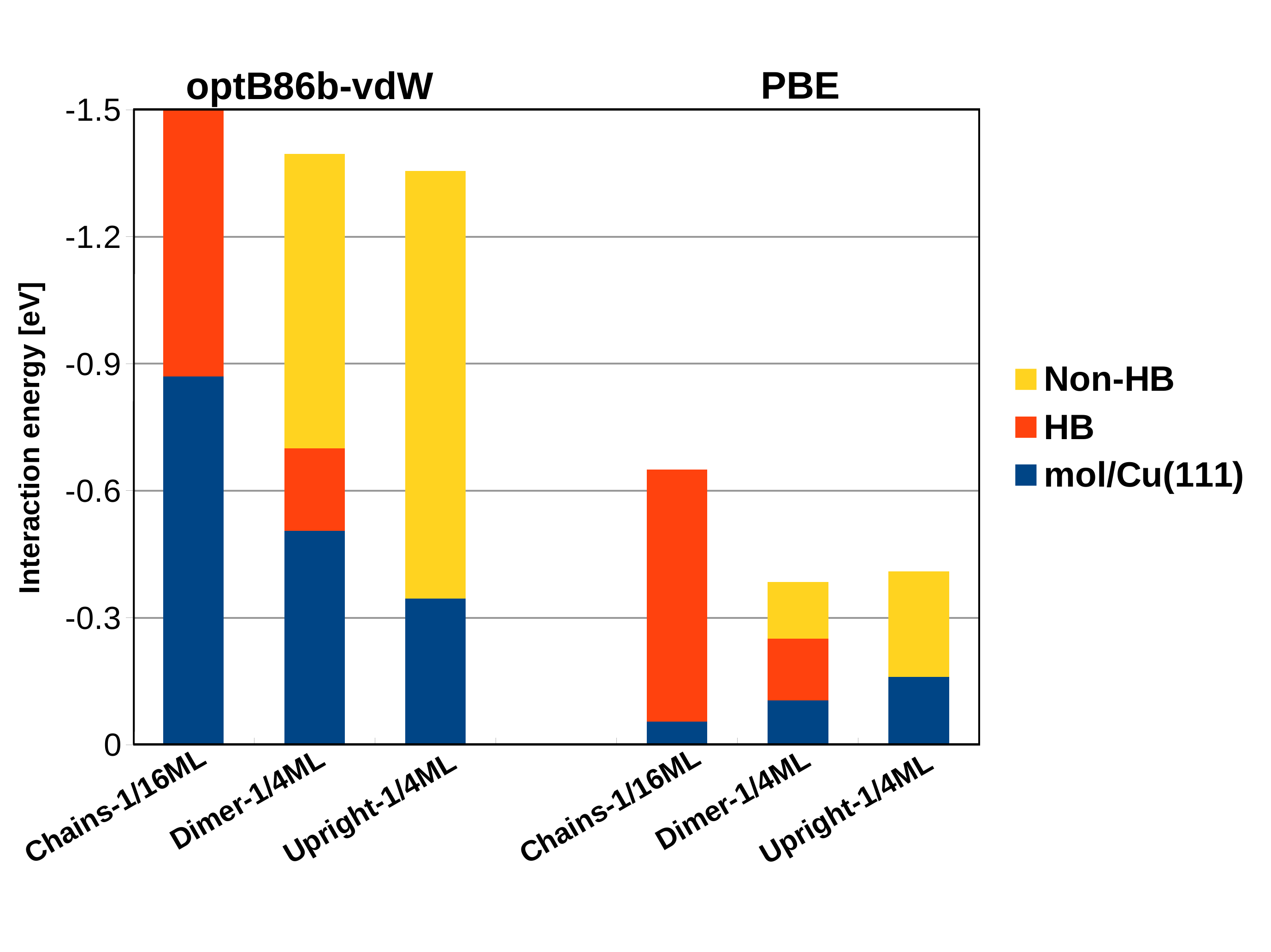}
 \caption{Interaction energies for low coverage chains, high coverage dimers and upright molecules. The contributions to the interaction energy are represented by different colours: in blue are the contributions from the interaction of the molecules with the surface, in red interactions from hydrogen bonds and in yellow intramolecular interactions not from HB (van der Waals and electrostatic interactions). The components of the decompositions were calculated using Eqs.~\ref{decomp1}-\ref{decomp3}.}
\label{fig:int_energies}
\end{figure}

In the range of pure physisorption, BTAH molecules form regular patterns in the form of hydrogen bonded chains with the molecules parallel to the surface with the N2 atom on the top copper site (Fig.~\ref{fig:str_cov}b).
The `Tilted' structure in Fig.~\ref{fig:alt_fig_1} has an adsorption energy $\sim 0.40$ eV/mol higher (less stable) than this HB configuration ($E_{\mathrm{ads}}=-1.46$ eV/mol).
This shows that the study of the interaction of isolated molecules on the surface gives only a limited description of the behaviour of the system and that the study of supramolecular aggregates is necessary to establish the full picture, even at low coverages.
The adsorption energy of the HB chain at 1/16 ML can be decomposed into a contribution of $\sim-0.92$ eV/mol from the physisorption of BTAH onto the copper surface, and a $\sim-0.59$ eV/mol contribution from hydrogen bonding (Fig~\ref{fig:int_energies}).
It is interesting to note that even with PBE, where the adsorption energy of a flat isolated molecule is very low, the flat hydrogen bonded structure is the most stable adsorption structure (with adsorption energy $ E_{\mathrm{ads}}=-0.61$ which is $\sim{0.20}$ eV lower than for the `Up' structure).
Fig.~\ref{fig:int_energies} shows that this is due to the strength of the hydrogen bond ($\mathrm{E}_{\mathrm{HB}}\sim{-0.51}$ eV/mol) which contributes to more than $90\%$ of the adsorption energy.

Hydrogen bonding dominates adsorption also at 1/8 ML, however the closer packing of the molecules makes steric repulsive interactions between the hydrogen atoms of the benzene-like rings relevant, and forces the system in a mixed flat/upright configuration (Fig.~\ref{fig:str_cov}c).
In these non-flat configurations, the highly-directional HBs become weaker ($\mathrm{E}_{\mathrm{HB}} \sim 0.3-0.5$ eV/mol according to the torsion) and other forms of interaction, such as N-Cu or $\pi$-$\pi$ bonding, become relevant.
The flat HB structure seen for 1/16 ML and 1/10 ML, is still among the low energy structures (Fig.~\ref{fig:str_cov}a), with the chains sliding on top of one another because of the space restrictions.
The $\pi$-bonds between the vertically stacked benzene-like rings add $\sim -0.10$ eV/mol to the total energy of the structure, however the interaction of the molecules with the substrate diminishes and the BTAH-Cu contribution reduces to $\sim-0.67$ eV/mol on average.
This results in an overall destabilisation of the HB chains, from $E_{\mathrm{ads}}=-1.46$ eV/mol at 1/16 ML to $E_{\mathrm{ads}}=-1.33$ eV/mol.

The same packing issues are relevant at 1/4 ML where the molecules adsorb in an upright configuration (a representative structure is shown in Fig.~\ref{fig:str_cov}d).
At this coverage, HB and non-HB configurations (in the form of HB dimers and upright molecules, respectively) coexist in the low-energy regime.
\begin{figure}[h]
\centering
 \includegraphics[width=1.0\textwidth]{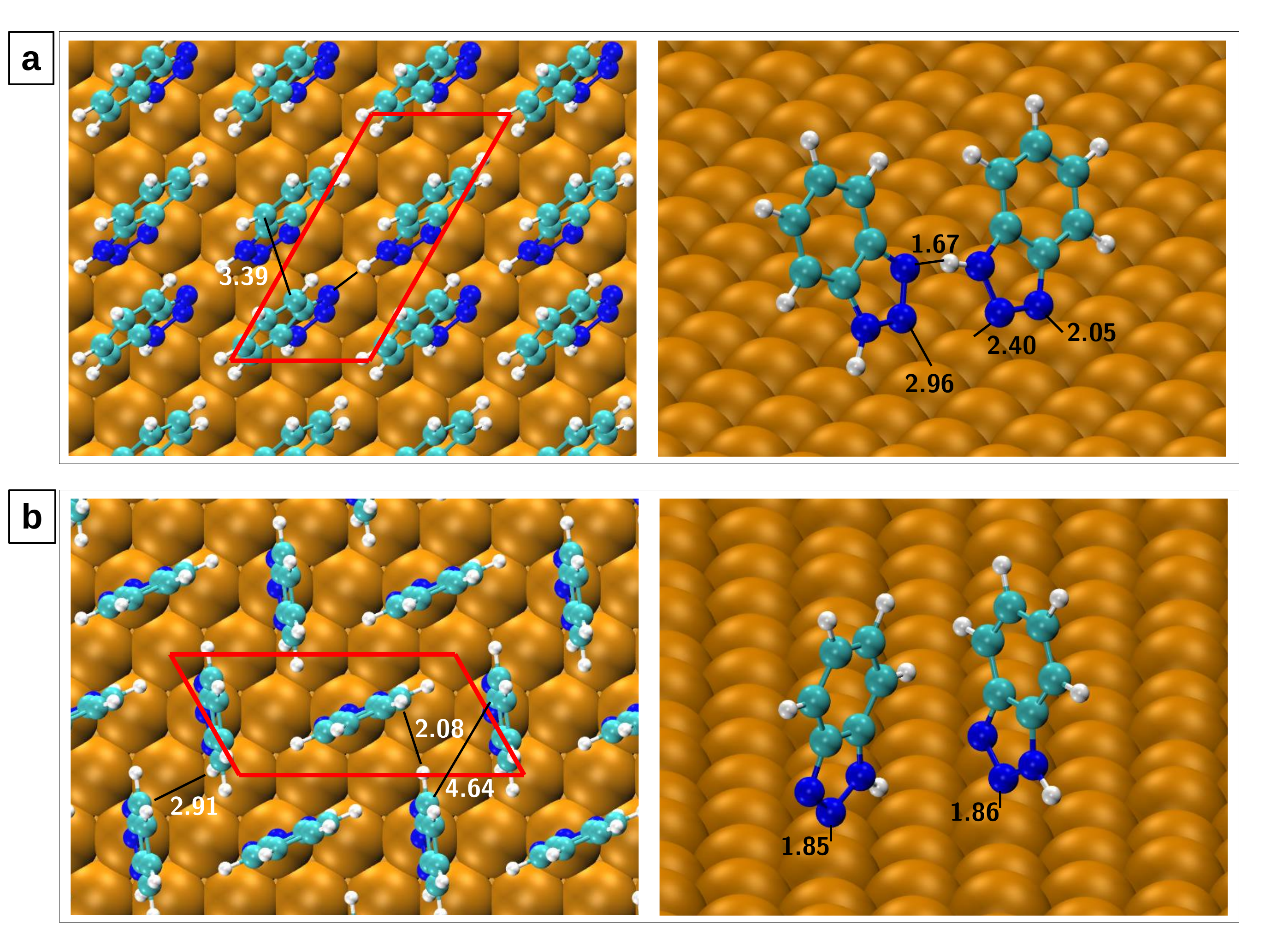}
 \caption{Top (left-hand-side) and side (right-hand-side) views of the most stable structures for BTAH at $1/4$ ML coverage a) HB dimer structure. b) Non-HB upright structure. The red box shows the unit cell. Selected distances are shown in \AA . In the top views the unit cell and periodic replicas are shown. In the side view only the molecules in the unit cell are shown for clarity.}
 \label{fig:str_cov2}
\end{figure}
In the HB structure (Fig.~\ref{fig:str_cov2}a), two upright BTAH molecules form a strong HB to the detriment of their interaction with the surface, with most N-Cu bonds much larger than the optimal length ($\sim 2.00$ \AA).
In the non-HB case (Fig.~\ref{fig:str_cov2}b), the BTAHs are preferentially packed at a $60^{\circ}$ angle with respect to one another, in a trade-off between the most stable anchoring of the upright molecules to the surface (with N2 and H10 on a copper top site) and avoiding lateral steric interactions between the H atoms in the benzene-like ring.
It can be seen in Fig.~\ref{fig:str_cov}a that, for $1/4$ ML, there are two HB structure with $\mathrm{E}_{\mathrm{ads}} \sim -1.00$ eV/mol and they both correspond to HB chains (with one HB per molecule).
The flat lying chain has already been discussed, while the chain of upright molecules owes its reduced stability to the distortion of the HBs whose strength for this configurations is only $\sim 0.1$ eV/mol.

Fig.~\ref{fig:int_energies} highlights the differences obtained when optimising the low- and high-coverage structures with optB86b-vdW or PBE.
It can be seen that high-coverage packing is less favourable with PBE than with optB86b-vdW.
The difference in energy between the low- and high-coverage structures is $\sim 0.3$ eV/mol with PBE, meaning that high-coverage configurations are higher in energy by almost $50\%$. 
On the other hand, the difference with optB86b-vdW is $\sim 0.2$ eV/mol, less than $15\%$ of the low-coverage interaction energy.
Indeed, at high-coverage, where HB becomes less important, intramolecular interactions (vdW and electrostatic), shown in yellow in Fig.~\ref{fig:int_energies}, become very important in stabilising the system.
In the case of PBE the contribution of vdW is not accounted for and the high coverage systems are thus less stable.

Overall we see that BTAH exhibits an incredibly rich phase behaviour when adsorbed on Cu(111), with hydrogen bonding dominating low energy adsorption before packing effects kick in and non-H-bonded configurations become competitive.
Moreover, significant differences between PBE and optB86b-vdW in the description of the adsorption behaviour of the molecules were seen for isolated molecules at low coverage and for high-coverage configurations.
Van der Waals forces increase the interaction between the molecules and the substrate and between the molecules in the overlayer, meaning that more strongly bound and higher coverage overlayers can be obtained.
This suggests that the use of vdW-inclusive functionals is necessary when studying the coverage dependence of molecular overlayers where intermolecular $\pi$ bonding is possible.

\section{BTA on Cu(111)}
\label{sub:bta_cu111}

As already mentioned, the study of protonated BTAH is likely to be relevant for acidic environments, whereas experimental evidence suggests that BTAH is deprotonated in alkaline solutions and in UHV~\cite{finsgar, walsh, grillo}.
It has been suggested that BTA forms organometallic structures on copper surfaces, \emph{i.e.} ordered networks of BTA molecules connected to one another via copper adatoms, \emph{in lieu} of the lost hydrogen atom which allows the formation of hydrogen bonded networks (see \emph{e.g.}~\cite{grillo, dugdale} and other references in~\cite{finsgar}).
Thus, understanding the structure of deprotonated benzotriazole on the Cu(111) surface is important from the point of view of applications (in alkaline environments) and as a direct link and validation with experimental work performed in vacuum conditions.

The most stable structures formed by BTA on Cu(111) are presented in Sec.~\ref{bta_1} and energetic considerations regarding the formation of the system, with respect to systems of fully protonated BTAH molecules are detailed in Sec.~\ref{bta_2}.
All calculations were performed with the optB86b-vdW functional.
\subsection{Structures and adsorption energies}
\label{bta_1}
\begin{figure}
\centering
 \includegraphics[width=1.0\textwidth]{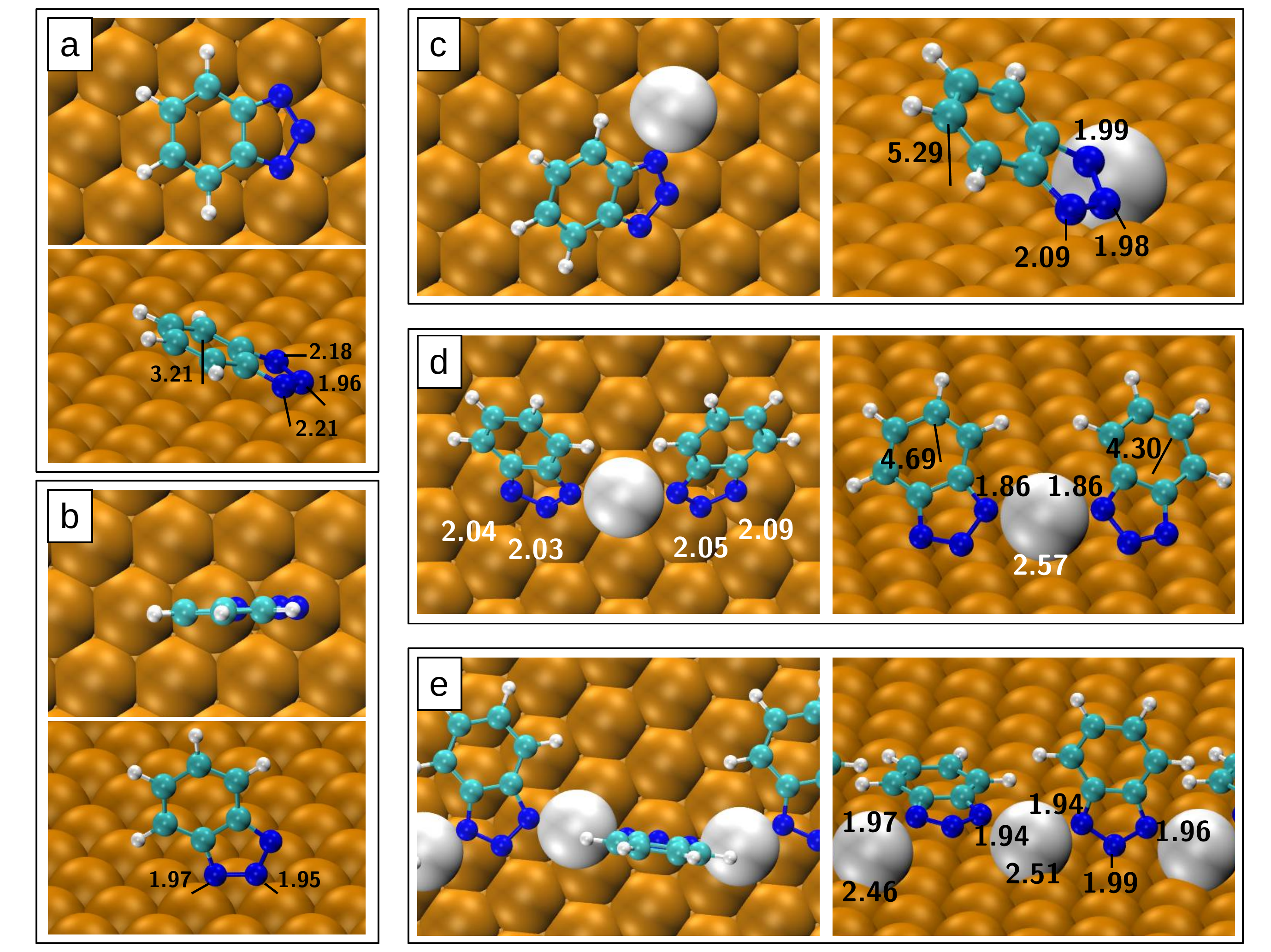}
 \caption{Top and side view of the most stable deprotonated BTA structures on Cu(111) at low coverage. The Cu adatom is shown in grey. a) Tilted configuration of BTA on Cu(111). b) Upright configuration of BTA on Cu(111). c) BTA-Cu$_{\mathrm{ad}}$ complex. The most stable position for the copper adatom was found to be the bridge site. d) BTA-Cu$_{\mathrm{ad}}$-BTA dimer. This is a representative structure of a low-energy configuration dimer with the copper adatom on the copper top site and a heavily tilted BTA. Other low-energy configurations include the copper adatom on the bridge site and upright BTA molecules. e) [BTA-Cu$_{\mathrm{ad}}$]$_n$ chain. The copper adatoms are found on the hollow sites. Selected distances are given in \AA\ and, for clarity, periodic replicas are not shown.}
 \label{fig:dep}
\end{figure}

\begin{table}
 \begin{tabular}{|c|c|}
   \hline
    Inhibitor & $E_{\mathrm{ads}}^{\mathrm{BTA}}$ \\
   \hline
   BTA (a,b) & -3.36 \\
   BTA-Cu$_{\mathrm{ad}}$ (c) & -3.94 \\
   BTA-Cu$_{\mathrm{ad}}$-BTA (d) & -4.02 \\
   BTA-Cu$_{\mathrm{ad}}$-BTA stacked & -4.24 \\
   $[\mathrm{BTA-Cu}_{\mathrm{ad}}]_{n}$ (e) & -4.55 \\
   \hline
   Corrosive atoms & $E_{\mathrm{ads}}$ \\
   \hline
   S & -4.79 \\
   Cl & -3.54 \\
   \hline
 \end{tabular}
 \caption{$E_{\mathrm{ads}}^{\mathrm{BTA}}$ for the BTA/Cu(111) structures shown in Fig.~\ref{fig:dep}. All values were calculated using Eq.~\ref{eq1} and are in eV/mol. Also reported are adsorption energies of two typical corrosive atoms, Cl and S. These adsorption energies (also in eV) are with respect to gas phase Cl and S atoms.}
 \label{table:bta}
\end{table}
The adsorption of single deprotonated molecules on Cu(111) and the formation of BTA-Cu$_{\mathrm{ad}}$ organometallic complexes on the surface were considered.
The most stable low-coverage (1/20-1/16 ML) structures are shown in Fig.~\ref{fig:dep}, for a single BTA (panels a and b), for a BTA-Cu$_{\mathrm{ad}}$ complex (panel c), for a BTA-Cu$_{\mathrm{ad}}$-BTA dimer (panel d) and for a [BTA-Cu$_{\mathrm{ad}}$]$_n$ chain (panel e).
Their adsorption energies (calculated using Eq.~\ref{eq1} with BTA as a reference) are shown in the top part of Table~\ref{table:bta}.
It can be seen that BTA interacts significantly more strongly ($> 2.0$ eV/molecule) with the surface than BTAH.
No appreciable charge transfer was seen between the H atoms and the BTA molecule when modelled in the same unit cell, suggesting the molecule undergoes a dehydrogenation process (rather than deprotonation).

For the isolated BTA, two degenerate low energy structures (E$_{\mathrm{ads}}=-3.36$ eV/mol) were found: a heavily tilted structure (Fig.~\ref{fig:dep}a), and an upright structure (Fig.~\ref{fig:dep}b).
In the tilted structure the bonding to the surface is obtained via all three nitrogen atoms (resulting in a $0.05$ \AA\ stretch in the N2N3 bond) and physisorption through the benzene-like ring.
Only the N2-Cu bond has the optimal length of $1.96$ \AA ; N2 and N3 form weaker bonds with a length $\sim 2.20$ \AA .
The upright structure is similar to the `Side' structure for BTAH, and the interaction with the surface occurs via two N-Cu bonds.\\
In the BTA-Cu$_{\mathrm{ad}}$ complex (Fig.~\ref{fig:dep}c, E$_{\mathrm{ads}}=-3.94$ eV/mol) the BTA molecule is tilted $\sim 70^{\circ}$ with respect to the surface and it forms three N-Cu bonds, two with the surface and one with the adatom (in the bridge site). 
Upright BTA-Cu$_{\mathrm{ad}}$ structures were found to have higher energy by $\sim{0.2}$ eV, although they were previously shown to be the most stable structure with PBE~\cite{peljhan_jpcc_2014}.\\
Thus, as with BTAH, we find that vdW forces alter the adsorption structures formed.
Isolated and stacked (with a distance of $\sim 4.4$ \AA\ between successive copper adatoms) dimers were investigated.
A number of almost degenerate low energy configurations (E$_{\mathrm{ads}}=-4.02$ eV/mol) were seen for the isolated dimer, in contrast with PBE results where upright dimers with Cu$_{\mathrm{ad}}$ at the top side was found to be the most stable structure. 
The inclinations of the molecule with respect to the surface in these structures was found to range from upright to $\sim 45^{\circ}$ inclination, and the copper adatom was found to sit in either the top or bridge position (a representative configuration is shown in Fig.~\ref{fig:dep}d).
In general, the lowest energy configurations were found when the distance between the nitrogen atom in the BTA molecule and copper adatom is not larger than d(N-Cu$_{\mathrm{ad}})\sim1.9$ \AA\ and the distance of Cu$_{\mathrm{ad}}$ with the substrate is not larger than d(Cu$_{\mathrm{ad}}$-$\mathrm{Cu(111))}\sim 2.5$ \AA.
In the case of stacked dimers, upright configurations were found to be preferred and the $\pi-\pi$ bonding between successive dimers stabilises the system by $\sim 0.2$ eV/mol.\\
The strongest adsorption energy, $E_{\mathrm{ads}}^{\mathrm{BTA}}=-4.55$ eV/mol, was found for the [BTA-Cu$_{\mathrm{ad}}$]$_n$ organometallic chains.
Fig.~\ref{fig:dep}e shows the most stable configuration, with the copper adatoms in hollow sites of the Cu(111) surface and one vertical BTA, forming three N-Cu bonds, two with the adatoms and one with the substrate, and a flat lying BTA, forming two N-Cu bonds with the two copper adatoms.
This structure is in good agreement with the structures found with PBE~\cite{peljhan_jpcc_2014, hakkinen} with however a higher rotation of the molecule which, in our case, lies almost flat while with PBE it is tilted by $\sim 60^{\circ}$ with respect to the surface.
Tests of stacked organometallic chains show that they do not significantly benefit from increased packing, yielding an adsorption energy which is degenerate with respect to the isolated chain.
Indeed, the flat-lying molecule in Fig.~\ref{fig:dep}e has to rotate at a $\sim 45^{\circ}$ angle with respect to the surface in order to allow for stacking, and the $\pi$ bonding with the surface is replaced with $\pi$ bonding with the benzene-like ring of the BTA in the parallel chains.\\
When the adsorption energies for the BTA systems are compared to the adsorption energy of atoms like sulphur or chlorine (bottom of Table~\ref{table:bta}), well-known active agents in the corrosion process of copper, it can be seen that their adsorption energies are very close, especially for chlorine.
Although this is a fairly crude comparison (corrosive agents in a realistic system are likely to be sulphur- or chlorine-containing molecules rather than isolated atoms), these values provide a hint that competitive adsorption could be an important factor in the corrosion inhibition process of BTA on copper.
\subsection{Energies of formation}
\label{bta_2}
\begin{figure}[h]
\centering
 \includegraphics[width=0.7\textwidth]{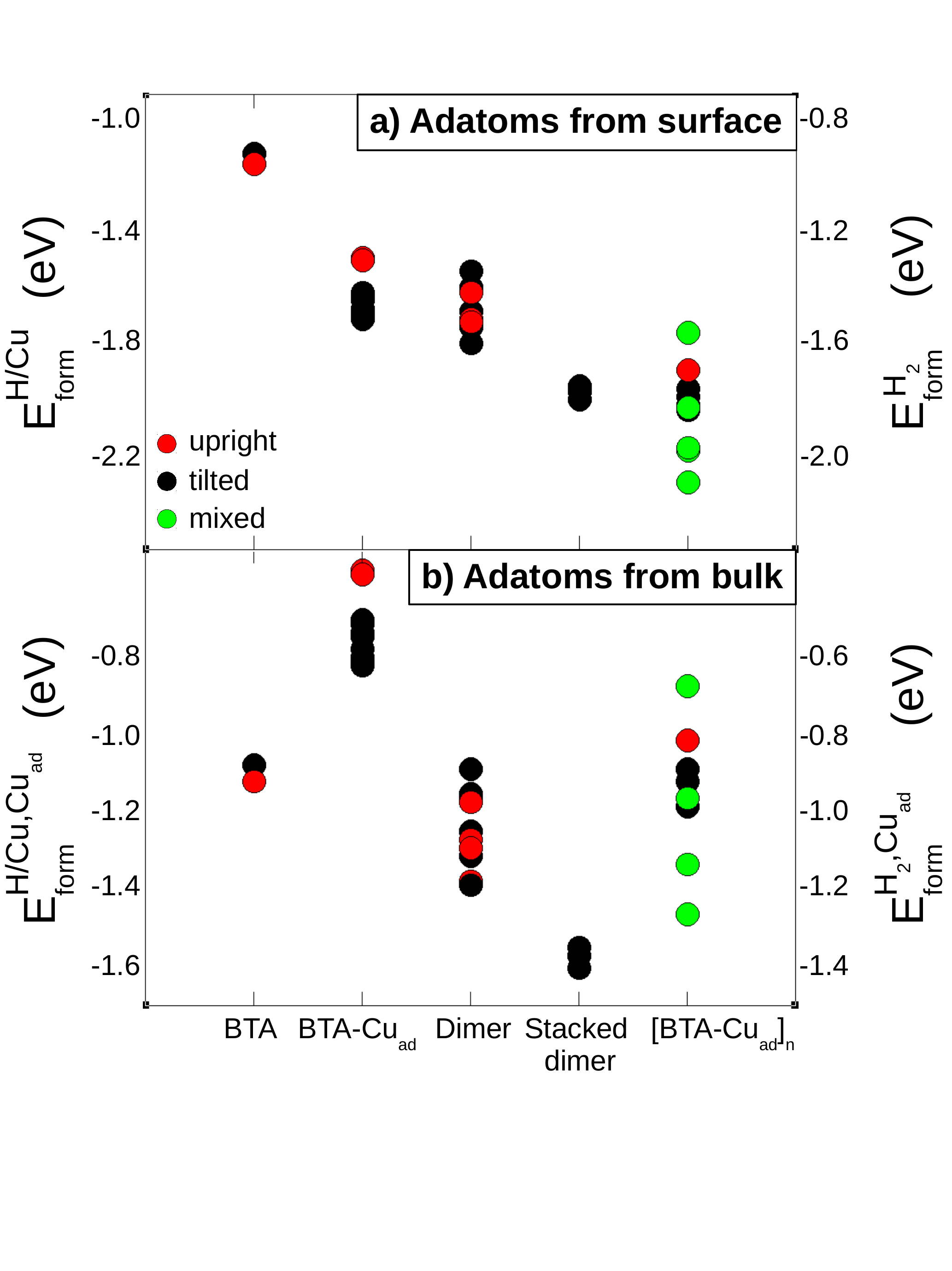}
 \caption{Formation energies for an isolated BTA, for a BTA-Cu$_{\mathrm{ad}}$ complex, for an isolated BTA-Cu$_{\mathrm{ad}}$-BTA dimer, for stacked dimers and for a [BTA-Cu$_{\mathrm{ad}}$]$_n$ necklace. The formation energies are calculated assuming that the H atom adsorbs on the surface (left-hand-side $y$-axis) or forms a gas-phase H$_2$ molecule (right-hand-side $y$-axis). a)~Formation energies for a BTAH dissociating and adsorbing on a surface where adatoms are already presents, following Eq.~\ref{reaction1}.  b)~Formation energies for a BTAH dissociating and adsorbing on a surface where adatoms are extracted from the bulk, following Eq.~\ref{reaction2} with $m=0$ for BTA, $m=1$ for BTA-Cu$_{\mathrm{ad}}$, dimer and stacked dimer, $m=2$ for [BTA-Cu$_{\mathrm{ad}}$]$_n$).}
 \label{fig:dep_adat}
\end{figure}

A direct comparison between the stability of the systems discussed in this and in Sec.~\ref{sub:btah_cu111} is given by their formation energies, calculated using Eqs.~\ref{reaction1}-\ref{reaction2} for BTA and E$_{\mathrm{form}}=$E$_{\mathrm{ads}}$ for BTAH (see Sec.~\ref{method}).
Four scenarios for the formation of the BTA systems are shown in Fig.~\ref{fig:dep_adat} (each point corresponds to a fully optimised BTA/Cu(111) structure).
In Fig.~\ref{fig:dep_adat}a formation energies according to Eq.~\ref{reaction1} are shown, \emph{i.e.} it is assumed that the surface contains defects and that copper adatoms are already present on the surface (in the hollow site) before the adsorption of the BTAH molecules.
The organometallic chains are the most stable system in this case.
In Fig.~\ref{fig:dep_adat}b formation energies according to Eq.~\ref{reaction2} are shown, \emph{i.e.} it is assumed that the surface is initially atomically flat and that the adatoms are extracted from the bulk.
The number of adatoms used in Eq.~\ref{reaction2} is $m=0$ for the BTA/Cu system, $m=1$ for the BTA-Cu$_{\mathrm{ad}}$ complex and for the BTA-Cu$_{\mathrm{ad}}$-BTA dimers and $m=2$ for the chains.
In this case the stacked BTA-Cu$_{\mathrm{ad}}$-BTA structure have the lowest energy, $\sim 0.2$ eV/mol more stable than the organometallic necklaces.
The two $y$ axes report energies for the two systems formed by the dissociated hydrogen atom: adsorbed on the copper surface (left-hand-side) and forming H$_2$ molecules (right-hand-side).

In Table~\ref{table:btah_form} the formation energies for the most stable structures in all four cases are reported.
\begin{table}[h]
 \begin{tabular}{|c|cc|}
   \hline
    Structure & $E_{\mathrm{form}}^{\mathrm{H/Cu(111)}}$ & $E_{\mathrm{form}}^{\mathrm{H}_2}$ \\
   \hline
   BTA (Fig.~\ref{fig:dep}a,b)  & -1.12 & -0.93\\
   BTA-Cu$_{\mathrm{ad}}$ (Fig.~\ref{fig:dep}c) & \cellcolor{cyan!20}-1.72 & \cellcolor{cyan!20}-1.51 \\
   BTA-Cu$_{\mathrm{ad}}$-BTA (Fig.~\ref{fig:dep}d)  & \cellcolor{cyan!20}-1.82 & \cellcolor{cyan!20}-1.61\\
   BTA-Cu$_{\mathrm{ad}}$-BTA stacked & \cellcolor{cyan!20}-2.02 & \cellcolor{cyan!20}-1.81\\
   $[\mathrm{BTA-Cu}_{\mathrm{ad}}]_{n}$ (Fig.~\ref{fig:dep}e) & \cellcolor{cyan!20}-2.31 & \cellcolor{cyan!20}-2.10 \\
   \hline
   Structure & $E_{\mathrm{form}}^{\mathrm{H/Cu(111), Cu}_{\mathrm{ad}}}$ & $E_{\mathrm{form}}^{\mathrm{H}_{2}\mathrm{,Cu}_{\mathrm{ad}}}$ \\
   \hline
   BTA (Fig.~\ref{fig:dep}a,b)  & -1.12 & -0.93\\
   BTA-Cu$_{\mathrm{ad}}$ (Fig.~\ref{fig:dep}c) & -0.86 & -0.65 \\
   BTA-Cu$_{\mathrm{ad}}$-BTA (Fig.~\ref{fig:dep}d) & -1.38 & -1.18\\
   BTA-Cu$_{\mathrm{ad}}$-BTA stacked & \cellcolor{cyan!20}-1.59 & -1.38 \\
   $[\mathrm{BTA-Cu}_{\mathrm{ad}}]_{n}$ (Fig.~\ref{fig:dep}e) & \cellcolor{cyan!20}-1.45 & -1.24 \\
   \hline
 \end{tabular}
\caption{$E_{\mathrm{form}}$ for the BTA/Cu(111) structures shown in Fig.~\ref{fig:dep}. The top part corresponds to the data shown in Fig.~\ref{fig:dep_adat}a. The values are calculated using Eq.~\ref{reaction1} \emph{i.e.} the formation energy of the adatoms is not taken into account. The bottom part corresponds to the data shown in Fig.~\ref{fig:dep_adat}b \emph{i.e.} it is assumed that the copper adatoms are extracted from the bulk (Eq.~\ref{reaction2}). The left column shows the formation energies when assuming that the dissociated H atoms adsorb on Cu(111), the right column when assuming they form gas phase H$_2$ molecules. All values are in eV/mol. The values highlighted in blue are those where E$_{\mathrm{form}}<-1.46$ eV, the formation energy of the BTAH flat HB structure.}
 \label{table:btah_form}
\end{table}
These are compared to the formation energy of the most stable BTAH/Cu(111) structure, the HB chains, and the blue highlighting indicates the deprotonated structures which are more stable than the BTAH/Cu(111) system (E$_{\mathrm{form}}<$E$_{\mathrm{form}}^{\mathrm{BTAH}}=-1.46$ eV).
When the formation energy of the copper adatom is not taken into account (top half of Table~\ref{table:btah_form}) most deprotonated structures, except for the isolated BTA molecule, are found to be more stable than the protonated HB chains.
If the formation energy of the copper adatom is considered (bottom half of Table~\ref{table:btah_form}), only the organometallic chains and the stacked dimers are found to be more stable that the BTAH HB chains, and only if the hydrogen atom is assumed to adsorb on the surface.
If the H atoms are assumed to associate into H$_2$ molecules the only competitive structure to the BTAH HB chains are the stacked dimers (E$_{\mathrm{form}}=-1.38$ eV/mol).
This shows that benzotriazole has a higher drive to deprotonate on defective surfaces with Cu adatoms than on atomically flat surfaces.

The results obtained from these four extreme cases can be used to give an approximate description of the behaviour of a real experimental system.
In the work of Grillo \emph{et al.}~\cite{grillo} the copper surface is reconstructed and therefore the mobility of the surface copper atoms is low thus making this system close to the case in Fig.~\ref{fig:dep_adat}b, \emph{i.e.} the copper adatoms need to be extracted from the bulk.
In this case (Table~\ref{table:btah_form}) the stacked BTA-Cu$_{\mathrm{ad}}$-BTA dimers, with the dissociated H atoms adsorbed on the surface, are the lowest energy configuration with E$_{\mathrm{form}}=-1.59$.
This is in excellent agreement with the STM experimental results, where stacked dimers are observed at low coverage and where the reconstruction of the Cu(111) surface, which is generally known not to reconstruct~\cite{crljen_prb_2003}, indicates the presence of impurities~\cite{MunozMarquez200597, moritani_jpcc_2008} (such as H atoms) on the surface.

\section{Discussion and conclusion}
\label{conclusions}

Results for the structures and energies of BTAH and BTA on Cu(111) have been presented here.
The adsorption of fully protonated BTAH is relevant to applications where the environment is acidic, whereas the adsorption of BTA on copper is relevant to alkaline environments and to compare with experiments of BTAH deposition on copper in vacuum conditions, where the molecule is found to deprotonate.
Benzotriazole is a complex molecule which requires treatment with a theoretical methodology capable of simultaneously describing chemisorption and physisorption.
In the present work DFT with a vdW-inclusive functional has been employed to optimise a large number of protonated and deprotonated structures on Cu(111).

We find that dispersion forces significantly alter the relative stabilities of adsorbed BTAH structures.
In addition to this, in the lowest energy tilted structure an interesting interplay with chemical bonding is found wherein dispersion forces bring the molecule close to the surface thus enhancing the chemical bonding of the molecule with the surface via the triazole group.
Whilst this is interesting and a potentially general effect, we have also shown that a single absorbed molecule provides limited insight into the behaviour of the overall system.
Indeed, while isolated BTAH preferentially adsorbs on copper via the azole nitrogen atoms and $\pi$-bonding of the carbon ring with the surface, the lowest energy structure at low coverage are HB chains of flat-lying molecules.

Overall we have found an incredibly rich coverage-dependent phase diagram for BTAH on Cu(111).
Three regimes were identified for the adsorption of BTAH on Cu(111), as a function of coverage: a low coverage hydrogen-bonded regime, where the molecules preferentially adsorb flat on the surface, an intermediate regime, where mixed flat and upright structures are observed, and a high coverage regime, where the molecules adsorb upright.
Steric interaction drive the change in configuration from the flat-lying physisorbed to the upright chemisorbed configuration of the BTAH.

The lowest energy configurations seen for BTA are either stacked BTA-Cu$_{\mathrm{ad}}$-BTA dimers or organometallic chains, according to whether the formation energy of the copper adatom is taken into account when calculating the stability of the complexes.
Good agreement is seen with experimental results in UHV, where the molecule was found to adsorb via the azole moiety in a vertical or near-vertical manner~\cite{fang_olson,walsh,grillo}.
In particular, the stacked dimer configuration was observed using STM to form on a reconstructed Cu(111) surface, where the mobility of the copper atoms is likely to be low and therefore copper adatoms require a large amount of energy to form.
In this case, the conditions of Fig.~\ref{fig:dep_adat}b apply, and indeed the expected configuration from calculations are stacked BTA-Cu$_{\mathrm{ad}}$-BTA dimers.

The use of a suitable exchange-correlation functional was found to be important for all the systems considered here.
Indeed, for an isolated BTAH on Cu(111) the `Flat' configuration was found to be favourable only when vdW dispersion forces were accounted for, and not in the case of PBE.
Moreover, the `Tilted' low coverage structure, which is the most stable with the optB86b-vdW functional and is favourable with all vdW functionals tested, is instead unstable when optimized with PBE (the `Flat' configuration is retrieved instead).
At high coverage, the lack of any description of $\pi$-$\pi$ bonding in PBE leads to a larger equilibrium distance between two BTAH molecules and thus to (comparatively) weaker adsorption energies for the high-coverage structures.
In the BTA/Cu systems it has been seen that PBE favours upright adsorption, because of the lack of dispersion interactions between the benzene-like ring and the surface.
When vdW interactions are accounted for a more complex behaviour is uncovered, especially for the BTA-Cu$_{\mathrm{ad}}$-BTA dimers where many degenerate low-energy structures are seen.

Both BTAH and BTA offer the possibility of forming fairly close packed layers on Cu(111) with little cost to the adsorption energy, and, for the case of BTA, with a gain in energy when the dimers are stacked, thus in principle offering a physical barrier to incoming corrosive molecules or atoms.
There is however a large difference ($\sim 2$ eV/mol) in the adsorption energy of the molecule with the surface between BTAH and BTA.
Since benzotriazole performs better as an inhibitor in alkaline conditions, where the likelihood of deprotonation is higher, there might be a link between the strongest interaction of the molecule with the surface and inhibition.
Indeed, the adsorption energy of BTA ($\sim4.00$ eV/atom) is fairly close to the adsorption energy of \emph{e.g.} two well known corrosive agents for copper.
It is likely that the actual corrosive agents in a corrosive solution are sulphur- or chlorine-containing molecules, rather than isolated atoms.
However, our calculations on chlorine and sulphur atoms give a ballpark estimate which suggests that competitive adsorption could be the key here for the success of benzotriazole as a corrosion inhibitor.

\section*{Acknowledgements}

A.M.'s work is partly supported by the European Research Council under the European Union’s Seventh Framework Programme (FP/2007-2013)/ERC Grant Agreement No. 616121 (HeteroIce project) and the Royal Society through a Wolfson Research Merit Award. 
The authors are grateful for computational resources to the London Centre for Nanotechnology and to the U.K. Car-Parrinello Consortium UKCP (EP/F036884/1), for access to HECToR.
C.G. would like to thank Dr. Federico Grillo for useful discussions.


%\balance

%If notes are included in your references you can change the title using the following command.

\footnotesize{
\bibliography{btah_cu} %your .bib file

\providecommand*{\mcitethebibliography}{\thebibliography}
\csname @ifundefined\endcsname{endmcitethebibliography}
{\let\endmcitethebibliography\endthebibliography}{}
\begin{mcitethebibliography}{43}
\providecommand*{\natexlab}[1]{#1}
\providecommand*{\mciteSetBstSublistMode}[1]{}
\providecommand*{\mciteSetBstMaxWidthForm}[2]{}
\providecommand*{\mciteBstWouldAddEndPuncttrue}
  {\def\EndOfBibitem{\unskip.}}
\providecommand*{\mciteBstWouldAddEndPunctfalse}
  {\let\EndOfBibitem\relax}
\providecommand*{\mciteSetBstMidEndSepPunct}[3]{}
\providecommand*{\mciteSetBstSublistLabelBeginEnd}[3]{}
\providecommand*{\EndOfBibitem}{}
\mciteSetBstSublistMode{f}
\mciteSetBstMaxWidthForm{subitem}
{(\emph{\alph{mcitesubitemcount}})}
\mciteSetBstSublistLabelBeginEnd{\mcitemaxwidthsubitemform\space}
{\relax}{\relax}

\bibitem[Breston({1952})]{breston}
J.~N. Breston, \emph{{{I}nd. and {E}ng. {C}hem.}}, {1952}, \textbf{{44}},
  {1755--1761}\relax
\mciteBstWouldAddEndPuncttrue
\mciteSetBstMidEndSepPunct{\mcitedefaultmidpunct}
{\mcitedefaultendpunct}{\mcitedefaultseppunct}\relax
\EndOfBibitem
\bibitem[Antonijevic and Petrovic(2008)]{antonijevic}
M.~M. Antonijevic and M.~B. Petrovic, \emph{{I}nt. {J}.~{E}lectrochem.\
  {S}ci.}, 2008, \textbf{3}, 1\relax
\mciteBstWouldAddEndPuncttrue
\mciteSetBstMidEndSepPunct{\mcitedefaultmidpunct}
{\mcitedefaultendpunct}{\mcitedefaultseppunct}\relax
\EndOfBibitem
\bibitem[{Procter \& Gamble, Ltd.}(1947)]{btah_patent}
{Procter \& Gamble, Ltd.}, \emph{British Patent 652339}, 1947\relax
\mciteBstWouldAddEndPuncttrue
\mciteSetBstMidEndSepPunct{\mcitedefaultmidpunct}
{\mcitedefaultendpunct}{\mcitedefaultseppunct}\relax
\EndOfBibitem
\bibitem[Finsgar and Milosev(2010)]{finsgar}
M.~Finsgar and I.~Milosev, \emph{{C}orr. {S}ci.}, 2010, \textbf{52}, 2737\relax
\mciteBstWouldAddEndPuncttrue
\mciteSetBstMidEndSepPunct{\mcitedefaultmidpunct}
{\mcitedefaultendpunct}{\mcitedefaultseppunct}\relax
\EndOfBibitem
\bibitem[Kokalj and Peljhan({2010})]{kokalj_lang}
A.~Kokalj and S.~Peljhan, \emph{{L}angmuir}, {2010}, \textbf{{26}},
  {14582}\relax
\mciteBstWouldAddEndPuncttrue
\mciteSetBstMidEndSepPunct{\mcitedefaultmidpunct}
{\mcitedefaultendpunct}{\mcitedefaultseppunct}\relax
\EndOfBibitem
\bibitem[Jiang and Adams(2003)]{jiang_2003}
Y.~Jiang and J.~B. Adams, \emph{{S}urf. {S}ci.}, 2003, \textbf{529}, 428\relax
\mciteBstWouldAddEndPuncttrue
\mciteSetBstMidEndSepPunct{\mcitedefaultmidpunct}
{\mcitedefaultendpunct}{\mcitedefaultseppunct}\relax
\EndOfBibitem
\bibitem[Peljhan and Kokalj({2011})]{peljhan_pccp_2011}
S.~Peljhan and A.~Kokalj, \emph{{{P}hys. {C}hem. {C}hem. {P}hys.}}, {2011},
  \textbf{{13}}, {20408--20417}\relax
\mciteBstWouldAddEndPuncttrue
\mciteSetBstMidEndSepPunct{\mcitedefaultmidpunct}
{\mcitedefaultendpunct}{\mcitedefaultseppunct}\relax
\EndOfBibitem
\bibitem[Peljhan \emph{et~al.}({2014})Peljhan, Koller, and
  Kokalj]{peljhan_jpcc_2014}
S.~Peljhan, J.~Koller and A.~Kokalj, \emph{{{J}.~{P}hys.\ {C}hem. C}}, {2014},
  \textbf{{118}}, {933--943}\relax
\mciteBstWouldAddEndPuncttrue
\mciteSetBstMidEndSepPunct{\mcitedefaultmidpunct}
{\mcitedefaultendpunct}{\mcitedefaultseppunct}\relax
\EndOfBibitem
\bibitem[Kokalj \emph{et~al.}(2010)Kokalj, Peljhan, Finsgar, and
  Milosev]{kokalj_jacs_2010}
A.~Kokalj, S.~Peljhan, M.~Finsgar and I.~Milosev, \emph{{J}.~{A}m.\ {C}hem.
  {S}oc.}, 2010, \textbf{132}, 16657\relax
\mciteBstWouldAddEndPuncttrue
\mciteSetBstMidEndSepPunct{\mcitedefaultmidpunct}
{\mcitedefaultendpunct}{\mcitedefaultseppunct}\relax
\EndOfBibitem
\bibitem[Chen and Hakkinen(2012)]{hakkinen}
X.~Chen and H.~Hakkinen, \emph{{J}.~{P}hys.\ {C}hem. C}, 2012, \textbf{116},
  22346\relax
\mciteBstWouldAddEndPuncttrue
\mciteSetBstMidEndSepPunct{\mcitedefaultmidpunct}
{\mcitedefaultendpunct}{\mcitedefaultseppunct}\relax
\EndOfBibitem
\bibitem[Dugdale and Cotton(1963)]{dugdale}
I.~Dugdale and J.~B. Cotton, \emph{{C}orr. {S}ci.}, 1963, \textbf{3}, 69\relax
\mciteBstWouldAddEndPuncttrue
\mciteSetBstMidEndSepPunct{\mcitedefaultmidpunct}
{\mcitedefaultendpunct}{\mcitedefaultseppunct}\relax
\EndOfBibitem
\bibitem[Roberts(1974)]{roberts_1974}
R.~F. Roberts, \emph{{J}.~{E}lectr.\ {S}pec. {R}el. {P}hen.}, 1974, \textbf{4},
  273\relax
\mciteBstWouldAddEndPuncttrue
\mciteSetBstMidEndSepPunct{\mcitedefaultmidpunct}
{\mcitedefaultendpunct}{\mcitedefaultseppunct}\relax
\EndOfBibitem
\bibitem[Rubim \emph{et~al.}(1983)Rubim, Gutz, Sala, and Orvillethomas]{rubim}
J.~C. Rubim, I.~G.~R. Gutz, O.~Sala and W.~J. Orvillethomas, \emph{{J}.~{M}ol.\
  {S}truct.}, 1983, \textbf{100}, 571\relax
\mciteBstWouldAddEndPuncttrue
\mciteSetBstMidEndSepPunct{\mcitedefaultmidpunct}
{\mcitedefaultendpunct}{\mcitedefaultseppunct}\relax
\EndOfBibitem
\bibitem[Rathgeber \emph{et~al.}({2012})Rathgeber, Bauer, Otto, Peter, and
  Wilde]{rathgeber}
S.~Rathgeber, R.~Bauer, A.~Otto, E.~Peter and J.~Wilde, \emph{{{M}icroelectr.
  {R}el.}}, {2012}, \textbf{{52}}, {2452--2456}\relax
\mciteBstWouldAddEndPuncttrue
\mciteSetBstMidEndSepPunct{\mcitedefaultmidpunct}
{\mcitedefaultendpunct}{\mcitedefaultseppunct}\relax
\EndOfBibitem
\bibitem[Fang \emph{et~al.}(1986)Fang, Olson, and Lynch]{fang_olson}
B.~S. Fang, C.~G. Olson and D.~W. Lynch, \emph{{S}urf. {S}ci.}, 1986,
  \textbf{176}, 476\relax
\mciteBstWouldAddEndPuncttrue
\mciteSetBstMidEndSepPunct{\mcitedefaultmidpunct}
{\mcitedefaultendpunct}{\mcitedefaultseppunct}\relax
\EndOfBibitem
\bibitem[Walsh \emph{et~al.}({1998})Walsh, Dhariwal, Gutierrez-Sosa, Finetti,
  Muryn, Brookes, Oldman, and Thornton]{walsh}
J.~F. Walsh, H.~S. Dhariwal, A.~Gutierrez-Sosa, P.~Finetti, C.~A. Muryn, N.~B.
  Brookes, R.~J. Oldman and G.~Thornton, \emph{{S}urf. {S}ci.}, {1998},
  \textbf{{415}}, {423}\relax
\mciteBstWouldAddEndPuncttrue
\mciteSetBstMidEndSepPunct{\mcitedefaultmidpunct}
{\mcitedefaultendpunct}{\mcitedefaultseppunct}\relax
\EndOfBibitem
\bibitem[Grillo \emph{et~al.}({2013})Grillo, Tee, Francis, Fruechtl, and
  Richardson]{grillo_nano_2013}
F.~Grillo, D.~W. Tee, S.~M. Francis, H.~A. Fruechtl and N.~V. Richardson,
  \emph{Nanoscale}, {2013}, \textbf{{5}}, {5269}\relax
\mciteBstWouldAddEndPuncttrue
\mciteSetBstMidEndSepPunct{\mcitedefaultmidpunct}
{\mcitedefaultendpunct}{\mcitedefaultseppunct}\relax
\EndOfBibitem
\bibitem[Grillo \emph{et~al.}({2014})Grillo, Tee, Francis, Fruechtl, and
  Richardson]{grillo}
F.~Grillo, D.~W. Tee, S.~M. Francis, H.~A. Fruechtl and N.~V. Richardson,
  \emph{{J}.~{P}hys.\ {C}hem. C}, {2014}, \textbf{{118}}, {8667}\relax
\mciteBstWouldAddEndPuncttrue
\mciteSetBstMidEndSepPunct{\mcitedefaultmidpunct}
{\mcitedefaultendpunct}{\mcitedefaultseppunct}\relax
\EndOfBibitem
\bibitem[Perdew \emph{et~al.}({1996})Perdew, Burke, and Ernzerhof]{pbe}
J.~P. Perdew, K.~Burke and M.~Ernzerhof, \emph{{{P}hys. {R}ev. {L}ett.}},
  {1996}, \textbf{{77}}, {3865}\relax
\mciteBstWouldAddEndPuncttrue
\mciteSetBstMidEndSepPunct{\mcitedefaultmidpunct}
{\mcitedefaultendpunct}{\mcitedefaultseppunct}\relax
\EndOfBibitem
\bibitem[Kresse and Hafner(1993)]{vasp_1}
G.~Kresse and J.~Hafner, \emph{{P}hys. {R}ev. B}, 1993, \textbf{47}, 558\relax
\mciteBstWouldAddEndPuncttrue
\mciteSetBstMidEndSepPunct{\mcitedefaultmidpunct}
{\mcitedefaultendpunct}{\mcitedefaultseppunct}\relax
\EndOfBibitem
\bibitem[Kresse and Hafner(1994)]{vasp_2}
G.~Kresse and J.~Hafner, \emph{{P}hys. {R}ev. B}, 1994, \textbf{49},
  14251\relax
\mciteBstWouldAddEndPuncttrue
\mciteSetBstMidEndSepPunct{\mcitedefaultmidpunct}
{\mcitedefaultendpunct}{\mcitedefaultseppunct}\relax
\EndOfBibitem
\bibitem[Kresse and Furthmueller(1996)]{vasp_3}
G.~Kresse and J.~Furthmueller, \emph{{C}omp. {M}at. {S}ci.}, 1996, \textbf{6},
  15\relax
\mciteBstWouldAddEndPuncttrue
\mciteSetBstMidEndSepPunct{\mcitedefaultmidpunct}
{\mcitedefaultendpunct}{\mcitedefaultseppunct}\relax
\EndOfBibitem
\bibitem[Kresse and Furthmueller(1996)]{vasp_4}
G.~Kresse and J.~Furthmueller, \emph{{P}hys. {R}ev. B}, 1996, \textbf{54},
  11169\relax
\mciteBstWouldAddEndPuncttrue
\mciteSetBstMidEndSepPunct{\mcitedefaultmidpunct}
{\mcitedefaultendpunct}{\mcitedefaultseppunct}\relax
\EndOfBibitem
\bibitem[Dion \emph{et~al.}({2004})Dion, Rydberg, Schroder, Langreth, and
  Lundqvist]{dion_prl_2004}
M.~Dion, H.~Rydberg, E.~Schroder, D.~C. Langreth and B.~I. Lundqvist,
  \emph{{{P}hys. {R}ev. {L}ett.}}, {2004}, \textbf{{92}}, {246401}\relax
\mciteBstWouldAddEndPuncttrue
\mciteSetBstMidEndSepPunct{\mcitedefaultmidpunct}
{\mcitedefaultendpunct}{\mcitedefaultseppunct}\relax
\EndOfBibitem
\bibitem[Carrasco \emph{et~al.}({2011})Carrasco, Santra, Klimes, and
  Michaelides]{carrasco_prl}
J.~Carrasco, B.~Santra, J.~Klimes and A.~Michaelides, \emph{{{P}hys. {R}ev.
  {L}ett.}}, {2011}, \textbf{{106}}, {026101}\relax
\mciteBstWouldAddEndPuncttrue
\mciteSetBstMidEndSepPunct{\mcitedefaultmidpunct}
{\mcitedefaultendpunct}{\mcitedefaultseppunct}\relax
\EndOfBibitem
\bibitem[Klimes and Michaelides({2012})]{klimes_review}
J.~Klimes and A.~Michaelides, \emph{{{J}.~{C}hem.\ {P}hys.}}, {2012},
  \textbf{{137}}, 120901\relax
\mciteBstWouldAddEndPuncttrue
\mciteSetBstMidEndSepPunct{\mcitedefaultmidpunct}
{\mcitedefaultendpunct}{\mcitedefaultseppunct}\relax
\EndOfBibitem
\bibitem[Carrasco \emph{et~al.}({2014})Carrasco, Liu, Michaelides, and
  Tkatchenko]{carrasco}
J.~Carrasco, W.~Liu, A.~Michaelides and A.~Tkatchenko, \emph{{{J}.~{C}hem.\
  {P}hys.}}, {2014}, \textbf{{140}}, {084704}\relax
\mciteBstWouldAddEndPuncttrue
\mciteSetBstMidEndSepPunct{\mcitedefaultmidpunct}
{\mcitedefaultendpunct}{\mcitedefaultseppunct}\relax
\EndOfBibitem
\bibitem[Klimes \emph{et~al.}(2011)Klimes, Bowler, and Michaelides]{klimes}
J.~Klimes, D.~R. Bowler and A.~Michaelides, \emph{{P}hys. {R}ev. B}, 2011,
  \textbf{83}, 195131\relax
\mciteBstWouldAddEndPuncttrue
\mciteSetBstMidEndSepPunct{\mcitedefaultmidpunct}
{\mcitedefaultendpunct}{\mcitedefaultseppunct}\relax
\EndOfBibitem
\bibitem[Grimme(2006)]{grimme}
S.~Grimme, \emph{J. Comp. Chem.}, 2006, \textbf{27}, 1787\relax
\mciteBstWouldAddEndPuncttrue
\mciteSetBstMidEndSepPunct{\mcitedefaultmidpunct}
{\mcitedefaultendpunct}{\mcitedefaultseppunct}\relax
\EndOfBibitem
\bibitem[Tkatchenko and Scheffler(2009)]{tkatchenko}
A.~Tkatchenko and M.~Scheffler, \emph{Phys. Rev. Lett.}, 2009, \textbf{102},
  073005\relax
\mciteBstWouldAddEndPuncttrue
\mciteSetBstMidEndSepPunct{\mcitedefaultmidpunct}
{\mcitedefaultendpunct}{\mcitedefaultseppunct}\relax
\EndOfBibitem
\bibitem[Haas \emph{et~al.}(2009)Haas, Tran, and Blaha]{haas}
P.~Haas, F.~Tran and P.~Blaha, \emph{Phys. Rev. B}, 2009, \textbf{79},
  085104\relax
\mciteBstWouldAddEndPuncttrue
\mciteSetBstMidEndSepPunct{\mcitedefaultmidpunct}
{\mcitedefaultendpunct}{\mcitedefaultseppunct}\relax
\EndOfBibitem
\bibitem[Blochl({1994})]{paw}
P.~E. Blochl, \emph{{{P}hys. {R}ev. B}}, {1994}, \textbf{{50}}, {17953}\relax
\mciteBstWouldAddEndPuncttrue
\mciteSetBstMidEndSepPunct{\mcitedefaultmidpunct}
{\mcitedefaultendpunct}{\mcitedefaultseppunct}\relax
\EndOfBibitem
\bibitem[Kresse and Joubert(1999)]{paw_vasp}
G.~Kresse and D.~Joubert, \emph{{{P}hys. {R}ev. B}}, 1999, \textbf{59},
  1758\relax
\mciteBstWouldAddEndPuncttrue
\mciteSetBstMidEndSepPunct{\mcitedefaultmidpunct}
{\mcitedefaultendpunct}{\mcitedefaultseppunct}\relax
\EndOfBibitem
\bibitem[Neugebauer and Scheffler(1992)]{neugebauer_prb_1992}
J.~Neugebauer and M.~Scheffler, \emph{{{P}hys. {R}ev. B}}, 1992, \textbf{46},
  16967\relax
\mciteBstWouldAddEndPuncttrue
\mciteSetBstMidEndSepPunct{\mcitedefaultmidpunct}
{\mcitedefaultendpunct}{\mcitedefaultseppunct}\relax
\EndOfBibitem
\bibitem[Makov and Payne(1995)]{makov_prb_1995}
G.~Makov and M.~C. Payne, \emph{{{P}hys. {R}ev. B}}, 1995, \textbf{51},
  4014\relax
\mciteBstWouldAddEndPuncttrue
\mciteSetBstMidEndSepPunct{\mcitedefaultmidpunct}
{\mcitedefaultendpunct}{\mcitedefaultseppunct}\relax
\EndOfBibitem
\bibitem[Michaelides \emph{et~al.}({2004})Michaelides, Alavi, and
  King]{michaelides_prb_2004}
A.~Michaelides, A.~Alavi and D.~A. King, \emph{{Phys. Rev. B}}, {2004},
  \textbf{{69}}, {113404}\relax
\mciteBstWouldAddEndPuncttrue
\mciteSetBstMidEndSepPunct{\mcitedefaultmidpunct}
{\mcitedefaultendpunct}{\mcitedefaultseppunct}\relax
\EndOfBibitem
\bibitem[Musiani \emph{et~al.}({1987})Musiani, Mengoli, Fleischmann, and
  Lowry]{musiani_jec_1987}
M.~M. Musiani, G.~Mengoli, M.~Fleischmann and R.~B. Lowry,
  \emph{{{J}.~{E}lectroanal.\ {C}hem.}}, {1987}, \textbf{{217}},
  {187--202}\relax
\mciteBstWouldAddEndPuncttrue
\mciteSetBstMidEndSepPunct{\mcitedefaultmidpunct}
{\mcitedefaultendpunct}{\mcitedefaultseppunct}\relax
\EndOfBibitem
\bibitem[Jiang and Adams({2003})]{jiang_ss}
Y.~Jiang and J.~B. Adams, \emph{{{S}urf. {S}ci.}}, {2003}, \textbf{{529}},
  {428}\relax
\mciteBstWouldAddEndPuncttrue
\mciteSetBstMidEndSepPunct{\mcitedefaultmidpunct}
{\mcitedefaultendpunct}{\mcitedefaultseppunct}\relax
\EndOfBibitem
\bibitem[Atodiresei \emph{et~al.}({2009})Atodiresei, Caciuc, Lazic, and
  Bluegel]{atodiresei}
N.~Atodiresei, V.~Caciuc, P.~Lazic and S.~Bluegel, \emph{{Phys. Rev. Lett.}},
  {2009}, \textbf{{102}}, {136809}\relax
\mciteBstWouldAddEndPuncttrue
\mciteSetBstMidEndSepPunct{\mcitedefaultmidpunct}
{\mcitedefaultendpunct}{\mcitedefaultseppunct}\relax
\EndOfBibitem
\bibitem[Liu \emph{et~al.}({2013})Liu, Filimonov, Carrasco, and
  Tkatchenko]{liu_nc_2013}
W.~Liu, S.~N. Filimonov, J.~Carrasco and A.~Tkatchenko, \emph{{Nat. Comm.}},
  {2013}, \textbf{{4}}, {2569}\relax
\mciteBstWouldAddEndPuncttrue
\mciteSetBstMidEndSepPunct{\mcitedefaultmidpunct}
{\mcitedefaultendpunct}{\mcitedefaultseppunct}\relax
\EndOfBibitem
\bibitem[Crljen \emph{et~al.}(2003)Crljen, Lazic, Sokcevic, and
  Brako]{crljen_prb_2003}
Z.~Crljen, P.~Lazic, D.~Sokcevic and R.~Brako, \emph{{Phys. Rev. B}}, 2003,
  \textbf{68}, {195411}\relax
\mciteBstWouldAddEndPuncttrue
\mciteSetBstMidEndSepPunct{\mcitedefaultmidpunct}
{\mcitedefaultendpunct}{\mcitedefaultseppunct}\relax
\EndOfBibitem
\bibitem[Muñoz-Márquez \emph{et~al.}(2005)Muñoz-Márquez, Parkinson, Quinn,
  Gladys, Tanner, Woodruff, Noakes, and Bailey]{MunozMarquez200597}
M.~A. Muñoz-Márquez, G.~S. Parkinson, P.~D. Quinn, M.~J. Gladys, R.~E.
  Tanner, D.~P. Woodruff, T.~C.~Q. Noakes and P.~Bailey, \emph{Surf. Sci.},
  2005, \textbf{582}, 97\relax
\mciteBstWouldAddEndPuncttrue
\mciteSetBstMidEndSepPunct{\mcitedefaultmidpunct}
{\mcitedefaultendpunct}{\mcitedefaultseppunct}\relax
\EndOfBibitem
\bibitem[Moritani \emph{et~al.}(2008)Moritani, Okada, Teraoka, Yoshigoe, and
  Kasai]{moritani_jpcc_2008}
K.~Moritani, M.~Okada, Y.~Teraoka, A.~Yoshigoe and T.~Kasai, \emph{J. Phys.
  Chem. C}, 2008, \textbf{112}, 8662\relax
\mciteBstWouldAddEndPuncttrue
\mciteSetBstMidEndSepPunct{\mcitedefaultmidpunct}
{\mcitedefaultendpunct}{\mcitedefaultseppunct}\relax
\EndOfBibitem
\end{mcitethebibliography}
\bibliographystyle{rsc}
}

\end{document}